\documentclass[acmsmall, screen]{acmart}
\usepackage{graphicx}
\usepackage{subcaption}
\usepackage{tabularx}
\usepackage{enumitem}
\usepackage[table]{xcolor}
\usepackage{makecell}
\usepackage{multirow}
\usepackage{array}
\usepackage{xcolor}

\AtBeginDocument{%
  }

\setcopyright{cc}
\setcctype{by-nc-nd}
\acmJournal{PACMHCI}
\acmYear{2026} \acmVolume{10} \acmNumber{6} \acmArticle{CSCW135}
\acmMonth{10} \acmDOI{10.1145/3816983}

\begin{document}

\title{PAPEL: A Collaborative System for Parental Guidance during Preschool Play-Based English Learning}

\author{Xutong Wang}
\email{wangxuto23@mails.tsinghua.edu.cn}
\orcid{0009-0009-6461-5446}
\affiliation{%
  \institution{Tsinghua University}
  \city{Beijing}
  \country{China}
}

\author{Yu Mei}
\orcid{0009-0008-3126-2974}
\email{meiy24@mails.tsinghua.edu.cn}
\affiliation{%
  \institution{Tsinghua University}
  \city{Beijing}
  \country{China}
}

\author{Qinwei Li}
\orcid{0009-0000-8132-9091}
\email{liqw24@mails.tsinghua.edu.cn}
\affiliation{%
  \institution{Tsinghua University}
  \city{Beijing}
  \country{China}
}

\author{Muyu Liu}
\orcid{0009-0007-2499-0984}
\email{liumuyu24@mails.tsinghua.edu.cn}
\affiliation{%
  \institution{Tsinghua University}
  \city{Beijing}
  \country{China}
}

\author{Xiwen Yao}
\orcid{0009-0001-5897-0312}
\email{xiiiwen.yao@gmail.com}
\affiliation{%
  \institution{Tsinghua University}
  \city{Beijing}
  \country{China}
}

\author{Chang Liu}
\orcid{0000-0002-1444-0993}
\email{c-liu21@tsinghua.org.cn}
\affiliation{%
  \institution{Tsinghua University}
  \city{Beijing}
  \country{China}
}

\author{Zhoutong Ye}
\orcid{0009-0009-9929-4734}
\email{yezt24@mails.tsinghua.edu.cn}
\affiliation{%
  \institution{Tsinghua University}
  \city{Beijing}
  \country{China}
}

\author{Jie Cai}
\authornote{Corresponding author.}
\orcid{0000-0002-0582-555X}
\email{jie-cai@mail.tsinghua.edu.cn}
\affiliation{%
  \institution{Tsinghua University}
  \city{Beijing}
  \country{China}
}

\author{Chun Yu}
\orcid{0000-0003-2591-7993}
\email{chunyu@tsinghua.edu.cn}
\affiliation{%
  \institution{Tsinghua University}
  \city{Beijing}
  \country{China}
}

\author{Yuanchun Shi}
\orcid{0000-0003-2273-6927}
\email{shiyc@tsinghua.edu.cn}
\affiliation{%
  \institution{Tsinghua University}
  \city{Beijing}
  \country{China}
}

\renewcommand{\shortauthors}{Wang et al.}

\begin{abstract}

Play-based parent-child interaction offers preschoolers rich opportunities for everyday foreign language learning, yet many parents struggle to turn open-ended play into effective English-as-a-Foreign-Language (EFL) learning experiences at home. To explore how AI might support this process, we conducted formative studies through interviews and a Wizard-of-Oz study. We identified four key challenges: content selection, language expression, balancing instruction and play, and problem solving.
To address these challenges, we present PAPEL, a parent–AI collaborative system that grounds suggestions in the ongoing play scene and organizes support into four core modules: content generation, language adaptation, balance assessment, and extended response. In a counterbalanced within-subjects study with 16 parent–child dyads, PAPEL was associated with more integrated parent utterances that combined playful and instructional content, as well as more parent–child conversational turns, than the lightweight chatbot baseline used in our study.



\end{abstract}

\ccsdesc[500]{Human-centered computing~Interactive systems and tools}
\ccsdesc[500]{Human-centered computing~Empirical studies in HCI}
\ccsdesc[300]{Applied computing~Computer-assisted instruction}
\ccsdesc[100]{Applied computing~Collaborative learning}
\keywords{Parent-AI collaboration, Play-based learning, Preschool English learning, AI-assisted early childhood education}

\begin{teaserfigure}
  \centering
  \includegraphics[trim=0 0 0 0, clip, width=1\linewidth]{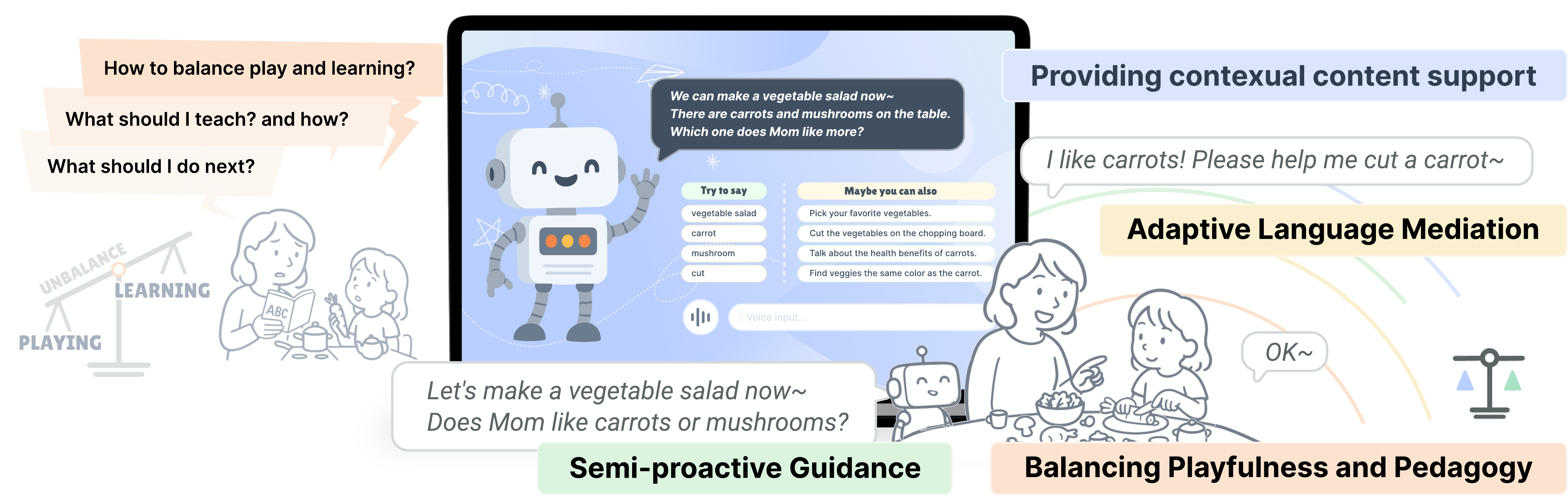}
  \caption{In a parent-child pretend-play scenario of making a vegetable salad, PAPEL helps parents weave English learning into the ongoing play without taking over the child’s playful experience. It suggests contextually relevant content, adapts language into child-friendly utterances, provides semi-proactive guidance to sustain the interaction, and helps balance playful engagement with pedagogical goals.}
  \label{fig:teaser}
\end{teaserfigure}

\received{May 13, 2025}
\received[revised]{January 13, 2026}
\received[accepted]{April 9, 2026}

\maketitle
\section{Introduction}

AI-assisted learning in home environments has become an increasingly important area of research in HCI and CSCW, particularly for preschool-aged children \cite{10.1145/3479585,10.1145/3686998,hiniker2016not,10.1145/3687018}. This unique developmental period is foundational for later cognitive, social, and linguistic skills, with early experiences in exploration and interaction strongly predicting future academic achievement \cite{ginsburg2007importance, ryan2000self, yogman2018power}. 

Theoretical frameworks like Parent-Mediated Theory (PMT)~\cite{singh2019kidsguard,connell2015parental} and Joint Media Engagement (JME)~\cite{yu2024parent} clearly highlight the critical role of parents in parent–child–AI interactions \cite{takeuchi2011new,10.1145/3772318.3790687,10.1145/3772318.3790479,10.1145/3772318.3791622}. These perspectives argue that meaningful media engagement does not happen when a child interacts with technology alone, but through active, collaborative engagement where the parent shapes the pace, content, and quality of the experience \cite{connell2015parental}. Consequently, this parent-centric principle has profoundly influenced the design of AI for in-home learning. Existing work positions AI as a supportive mediator in a triadic interaction, offering resources and scaffolding to empower parents \cite{ho2024s,ho2023designing,10.1145/3772318.3791970,10.1145/3715336.3735837}. Such a position establishes the parent's central role, tasking them with guiding the learning process while responding to their child’s needs.

Current AI systems for home-based learning have largely focused on structured, goal-directed activities, such as co-reading \cite{xu2023rosita, Social_robots_as_conversational_catalysts,home-study-of-co-reading}, storytelling \cite{zhang2022storybuddy, Storytelling, mathemyths}, and game-based learning \cite{xu2023mathkingdom}. However, open-ended play-based learning remains underexplored. In our work, we define ``play'' as the process in which children use their imagination to engage in a series of interactions while playing with toys, not limited to pretend play. When this process is guided by educational objectives, it constitutes play-based learning. In such scenarios, parents must navigate multiple challenges, including clarifying learning goals, fostering rich language interactions, and sustaining children’s interest~\cite{diederich2024changing, li2024parent, cheng2014children}. 
These challenges require strong pedagogical sensitivity and adaptability—skills that can be difficult for parents without formal training~\cite{pedagogical_play, pedagogical-play-types}. More importantly, in open-ended play, parents must continuously shift between \textbf{multiple roles}, such as playmate and educator~\cite{lin2020parents}, while responding to the child’s evolving interests. This makes play-based learning a particularly demanding setting for parent-centered AI support.


In this study, we explore how AI can help to facilitate EFL (English as a Foreign Language) learning through play from parents' perspectives. EFL is highly valued in East Asian families \cite{butler2015english,provsic2020english}, but due to parents' own lack of confidence or capability, they struggle to create an effective English learning environment at home \cite{on2024examining}. 

To systematically understand parents' challenges in AI-assisted, play-based English learning for
preschoolers, we conducted a two-phase formative study including parent and teacher interviews and a Wizard-of-Oz study. This process revealed four key challenges in open-ended scenarios: (1) Parents struggle to find content suited to their child’s level and play context, (2) Parents fail to express and deliver knowledge clearly, (3) Parents struggle to balance play and learning during interactions, and (4) Parents lack the ability to prompt the AI to ensure the process stays on track.
To address these challenges, we designed and developed PAPEL (\textbf{P}arent–\textbf{A}I collaborative system for \textbf{P}lay-based \textbf{E}nglish \textbf{L}earning), a multimodal large language model (MLLM)-driven, conversational multi-agent system that joins parent–child interactions and provides in-situ support with minimal prompting burden on parents.
PAPEL includes four core modules, each targeting one challenge: (C1) Content Generation suggests appropriate content, (C2) Language Adaptation adjusts response style, (C3) Balance Assessment monitors balance between play and learning, and (C4) Extended Response proactively expands responses to nudge the situation. 

To assess the system, we conducted a counterbalanced user study involving 16 parent-child dyads, comparing PAPEL with a baseline system featuring a standard LLM chatbot. Results suggest that parents became better at turning the ongoing play scene into teachable activities, articulating English more fluently and appropriately, keeping play and instruction in better balance, and steering the interaction forward when it began to stall. To interpret these findings, we draw on Role Theory~\cite{role_theory} to understand tensions in parents’ shifting roles during play-based learning, and on the Pedagogical Play Framework~\cite{pedagogical_play} to discuss how AI can better support pedagogical work in open-ended family play.

In summary, this study contributes the following:

\textbullet\  We present PAPEL, a parent-centric AI system designed to facilitate play-based English learning between parents and children.

\textbullet\   Through a formative study involving interviews and a Wizard-of-Oz experiment, we analyze the difficulties associated with incorporating AI-assisted English language instruction into a play-based learning environment.
Through a counterbalanced user study, we demonstrate that PAPEL effectively supports parents in play-based learning, and reveal parents’ role conflict between “playmate” and “educator” roles.

\textbullet\ 
Building directly on the user-study findings, we apply Role Theory and the Pedagogical Play Framework to explain why the tension emerges in play-based learning and translate it into concrete design implications for future AI systems.



\section{Related Work}
\subsection{Parents in Play-based Learning}
Children naturally enjoy using toys to engage their imagination and enact a series of interactions; when these activities are integrated with educational goals, they constitute play-based learning~\cite{taylor2020play}.
Play-based learning holds an irreplaceable role in home education~\cite{bubikova2019ece}, as ``play'' typically involves voluntary engagement, minimal external control, and a strong sense of autonomy \cite{early2009behavioural}. It aligns with children's natural inclination for hands-on activities and provides children with an environment to explore and express themselves freely \cite{ryan2000self, yogman2018power}. The multimodal nature of play enhances immersion and memory retention \cite{hirsh2004einstein, ciolan2013play, cai2023case, hirsh2009mandate}. 
However, despite its flexible and child-centered nature, play-based learning still requires adult involvement to fully realize its educational potential \cite{hughes2009children,moyles2014excellence}.

According to \textit{Pedagogical Play Framework} \cite{pedagogical_play}, effective play-based learning should be built upon purposeful involvement, where educators dynamically adjust between guiding learning goals and supporting children's active engagement. This implies that effective play-based learning at home also requires parental involvement \cite{nagle2021pathfinder}.
Effective parental observation and guidance significantly enhance children's cognitive development \cite{li2024parent, cheng2014children}, helping them grasp complex concepts \cite{budiyanto2021developing}, foster creativity, and develop problem-solving skills \cite{druga2022families}. Language learning particularly benefits from familiar adult interaction, increasing children's language production and targeted feedback reception \cite{diederich2024changing}.

However, parents face challenges in teaching skills during role-play \cite{park2024limitations}, balancing digital and traditional tools \cite{hiniker2018let}, and navigating generational communication barriers \cite{druga2022families}. These limitations constrain home-based play's educational potential and highlight the need for context-sensitive parental support mechanisms.

Therefore, AI technologies may offer opportunities to fill the gap.

\subsection{Parent-Supporting AI in Family Education}
\label{rw:AI in family education}
Prior work has largely framed AI use in family learning through parent-centric lenses including Parent-Mediated Theory~\cite{singh2019kidsguard,connell2015parental} and Joint Media Engagement~\cite{yu2024parent}, which together motivate a “parent-in-the-loop” paradigm: parents are expected to shape how children engage with AI at home \cite{ho2024s,cagiltay2024toward,chen2022designing,ChatGPT-Use,yu2021parental_facilitation}.
Building on this paradigm, recent HCI/CSCW AI systems mainly support parents in structured, goal-directed activities, where the parent takes a leading role~\cite{10.1145/3462204.3481759,10.1145/3637409}. In co-reading \cite{Social_robots_as_conversational_catalysts,xu2023rosita,home-study-of-co-reading,dietz2024contextq} and digital storytelling \cite{mathemyths,Storytelling,zhang2022storybuddy,lee2024open,chen2024storysparkqa}, AI scaffolds parent-led dialogue and turn-taking to help children develop vocabulary and narrative understanding. 
In game-based learning \cite{xu2023mathkingdom}, parents typically act as instructors and playmates---explaining rules, steering toward predefined goals, and practicing targeted skills \cite{sharma2016promoting,long2022family,coo}. Beyond these task-focused settings, some systems also foreground parents’ caregiving and relational roles by directly facilitating parent–child communication and more natural interaction \cite{choi2024aacesstalk,shi2025need,ho2023designing,chan2017wakey}. 
Taken together, existing work has begun to support parents’ multiple roles---both instructional and caregiving---but largely within structured scenarios where goals are pre-specified and easier to manage.

In contrast, play-based learning remains less explored. It is characterized by child-led exploration, where interaction goals are dynamic and continually reshaped by children’s moment-to-moment curiosity \cite{ryan2000self,yogman2018power}. While there is preliminary work in toy-based interaction \cite{kwon2022captivate,liu2025bricksmart,kozima2023communication,gvirsman2020patricc}, these efforts often rely on predefined content or constrained implementations, offering limited support for parents as they navigate and switch between roles in open-ended play.

\subsection{Research Context}
Our work focuses on preschoolers. Typically aged 3 to 6, preschoolers are in a stage of rapid development \cite{ginsburg2007importance, ryan2000self, yogman2018power}. They are naturally curious and learn best through play and hands-on experiences \cite{kalogiannakis2018teaching}. Their rich imagination and creativity flourish during this period, making parental support crucial for guiding and scaffolding their play \cite{nagle2021pathfinder}. Preschool is also a particularly suitable period for introducing a new language, as young children are especially responsive to linguistic input during this stage, and research on early bilingualism suggests that dual-language development can proceed successfully in the early years \cite{lockiewicz2018early,byers2013bilingualism}.

A profound cultural emphasis on early education in many East Asian households places a high priority on English from as early as preschool \cite{butler2015english,provsic2020english}. However, this ambition often creates a difficult learning environment for preschoolers, who face a significant gap between structured classroom lessons and the lack of authentic, playful opportunities to use English at home \cite{jin2018early}. This challenge is compounded when highly motivated parents lack the confidence or ability to create these playful scenarios \cite{on2024examining}. As a result, in-home English learning often becomes a source of pressure, feeling awkward and difficult to sustain for preschoolers in East Asia \cite{lan2011learning,hirsh2015contribution}.

Play-based learning is exceptionally well-suited to bridge this gap \cite{guz2016learning}. Language acquisition thrives on the immersion and situational need that play naturally provides. It transforms abstract language into tangible scenarios through play \cite{dorathy2011second}, embeds communication as a core game mechanic \cite{vives2018critical}, and allows children to immerse themselves in a foreign language within activities that align with their natural interests, fostering practical skill development \cite{piker2013understanding}.

Therefore, the intersection of high parental motivation for early EFL in East Asia and the inherent suitability of play-based learning for language acquisition creates a natural and compelling research context for our work with preschoolers.

\section{Formative Study}
\begin{figure}
    \centering
    \includegraphics[width=1\linewidth]{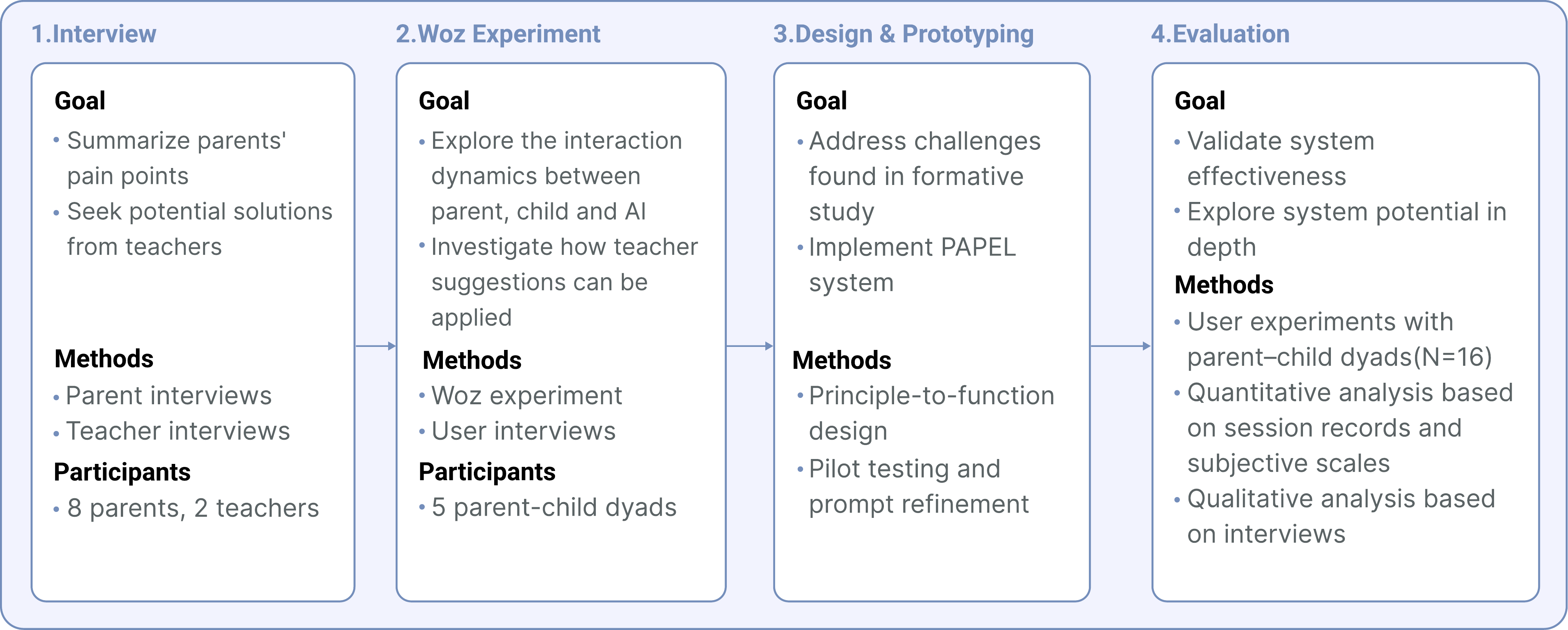}
    \caption{Study Procedure Overview.}
    \Description{Overview of the study procedure, as depicted in the figure.}
    \label{fig:design process}
\end{figure}

To design our system, we followed a user-centered research process consisting of multiple stages. First, we interviewed parents to understand the challenges they face when using pretend play to support their children's language learning. We then interviewed early childhood educators to gather expert-informed strategies for addressing these challenges. Finally, we conducted a Wizard-of-Oz (WoZ) experiment to validate and adapt these strategies into in-home scenarios and observe how parents would utilize AI support in practice. Across all stages, including the user study, we recruited participants through REDnote~\footnote{\url{https://www.xiaohongshu.com}}, a widely used online community platform in China where users actively share and discuss everyday topics, including parenting and early childhood education. This made it a practical channel for reaching parents likely to have experience with home learning and play activities. We used a convenience sampling approach with eligibility screening. For the formative interviews, we included parents with prior in-home play-based learning experience; for the user study, we used quota-oriented recruitment to help balance dyads by child age and prior in-home play-based English-learning experience.
The overall research process is illustrated in Fig.~\ref{fig:design process}.

\subsection{Parent Interview}
We conducted one-hour-long, one-on-one interviews with a group of 8 parents, most of whom held at least a bachelor’s degree (five bachelor’s, two master’s, one doctorate). Parents’ English ability spanned from limited to near-fluent, based on their own assessment. Their children were between 3 and 6 years old, with formal English learning experience ranging from none to 36+ months. All parents reported having previously taught English to their child at home. The interviews were conducted online, and the primary aim was to explore their motivations, challenges, and expectations regarding co-playing with their children. 

To ground our design in everyday home play, we asked parents what toys/materials they typically use for play-based learning at home. Parents most frequently mentioned kitchen sets, doctor kits, cars, and blocks. These categories informed the materials used in our study tasks. Our interviews also revealed that sustaining such play-based English learning at home entails several practical challenges for parents. Notably, parents primarily described play-based learning through physical toys and face-to-face interaction. A recurring concern was preschoolers’ screen exposure: several parents expressed reluctance to place phones or tablets at the center of the activity, which helps explain why non-digital play was foregrounded in their accounts. This concern also informed our focus on AI as a parent-facing support tool that augments in-person play, rather than as a child-facing screen activity. 

\subsubsection{Language Scaffolding Difficulties}

Parents often struggle to model correct English during pretend play due to limited vocabulary and grammar. Some avoid using English altogether, fearing they might teach their children incorrect usage. Even those with higher proficiency feel uncertain about which expressions are appropriate or effective in various game contexts. This lack of confidence leads to frequent code-switching and minimal English input during play. For children, parents’ inconsistent language input can also cause confusion.

\subsubsection{Improvisational Challenge in Gameplay}
The open-ended nature of play imposes a dual demand on parents: they must act as both a proactive narrator and a responsive playmate. They are responsible for scaffolding the narrative to sustain child engagement, yet must simultaneously improvise in response to their child's unpredictable ideas. Combining these roles is cognitively exhausting, and without clear prompts, parents often struggle to manage the interaction flow, leading to stalled play and feelings of inadequacy.

\subsubsection{Lack of Pedagogical Strategies}
Parents often wish to integrate educational goals into play, but many lack the pedagogical strategies to do so effectively. Without clear methods, their attempts to introduce learning moments—such as vocabulary or moral lessons—can become disjointed or intrusive, making the interaction feel awkward and breaking the natural flow of play. Children may be confused or even annoyed with too much instructive content.

\subsection{Teacher Interview}
To explore effective ways to support parents in play-based learning, we conducted one-to-one interviews with two kindergarten English teachers, each with more than five years of teaching experience. In our context, kindergarten English lessons are also organized around games, songs, and pretend-play activities instead of formal lecturing, so these teachers offered relevant expertise on designing playful English learning that can be adapted to parent–child interaction at home. Each interview, lasting around one hour, was conducted in different formats—one remotely via video call, and the other face-to-face.

T1 and T2 provided methods to help parents with vocabulary, language use, gameplay, and pedagogy. For vocabulary, T1 recommended starting with simple, high-frequency words and speaking English throughout, while T2 suggested a bilingual approach, introducing words through gestures and contextual cues. T1 also emphasized structured narratives, while T2 advocated for flexible, child-led gameplay and real-time adaptation. In terms of pedagogy, T1 suggested using pre-designed lesson plans, while T2 focused on adapting the content to children's interests.

Comparing the parent and teacher interviews, we found that the first two challenges identified by parents, difficulty coming up with appropriate content and difficulty expressing it during play, were broadly aligned with the kinds of support both teachers described. T1 emphasized structured lesson planning, whereas T2 stressed flexible, child-led adaptation. Despite this difference, both perspectives pointed to forms of support that current MLLMs may help scaffold, such as generating contextual vocabulary, suggesting possible activity progressions, and offering alternative ways of phrasing instructional language in real time. However, the third challenge, parents’ lack of pedagogical strategies for integrating play and instruction, appeared difficult to address through content support alone.
T1 and T2 emphasized that effectively fusing play and instruction requires a level of pedagogical sensitivity that many parents currently lack. This indicates that the gap lies not merely in content, but in the parental role itself, meaning that simple content generation is unlikely to suffice. This deeper, role-based challenge prompted us to conduct an in-depth study on how AI could meaningfully support parents in overcoming this pedagogical barrier.

\subsection{Wizard-of-Oz Experiment}
Informed by the parent and teacher interviews, we identified two broad directions of support: helping parents with content generation and language expression, and helping them manage the more difficult pedagogical challenge of integrating play and instruction. We therefore used the WoZ study to examine how these forms of support played out in situated parent-child-AI interaction and what additional needs emerged in practice. Specifically, our WoZ study focused on two questions:

\begin{itemize}
    \item 1. To what extent can AI support parents’ content generation and language expression during open-ended play-based English learning?
    \item 2. What tensions and unmet support needs emerge when parents try to integrate playful interaction and instructional goals using AI support?
\end{itemize}

\begin{figure}[H]
  \centering
  \begin{subfigure}[b]{0.46\textwidth}
    \includegraphics[width=\linewidth]{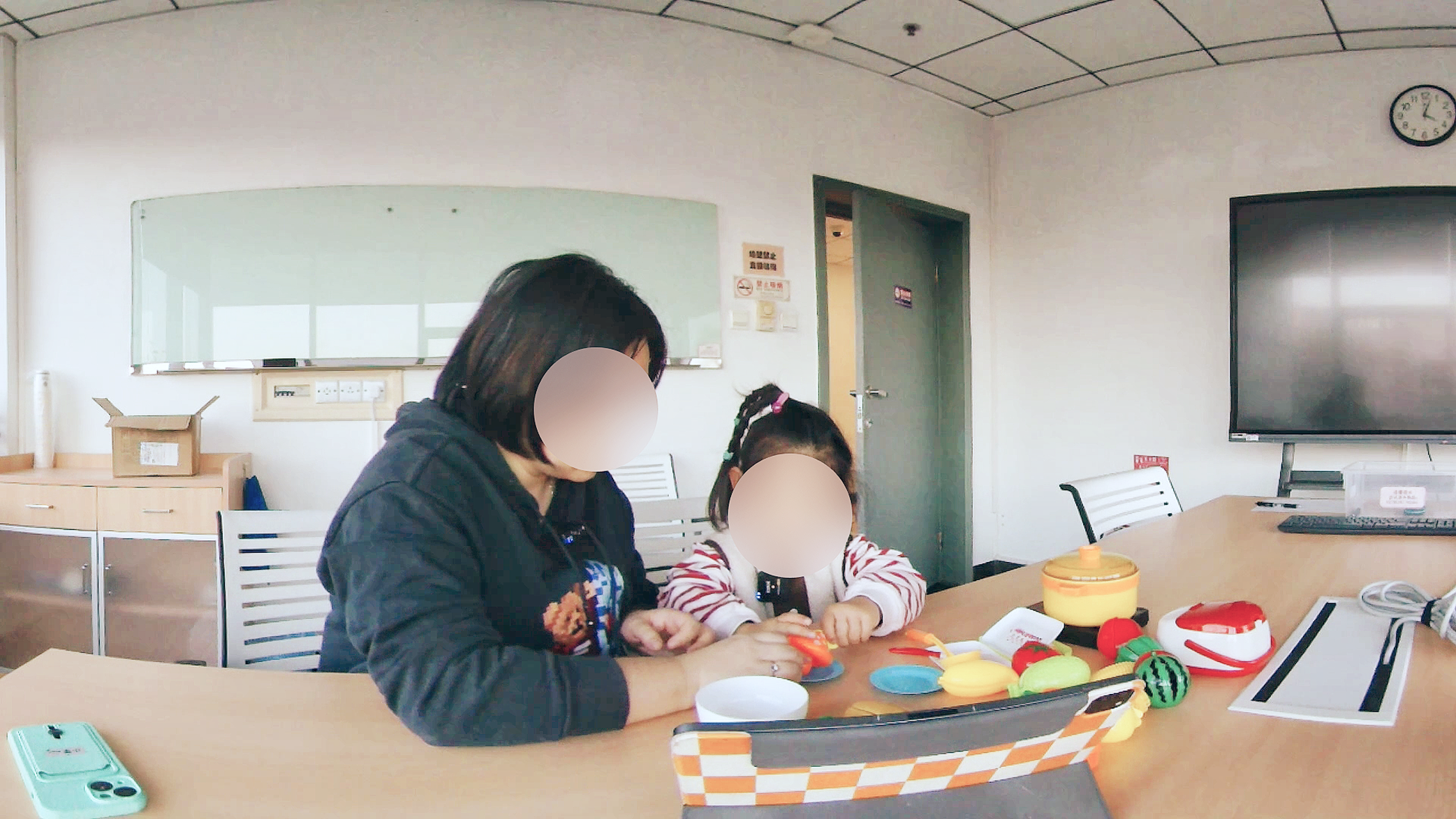}
    \caption{}
    \label{subfig:WoZ-setting}
  \end{subfigure}
  \begin{subfigure}[b]{0.44\textwidth}
    \includegraphics[width=\linewidth]{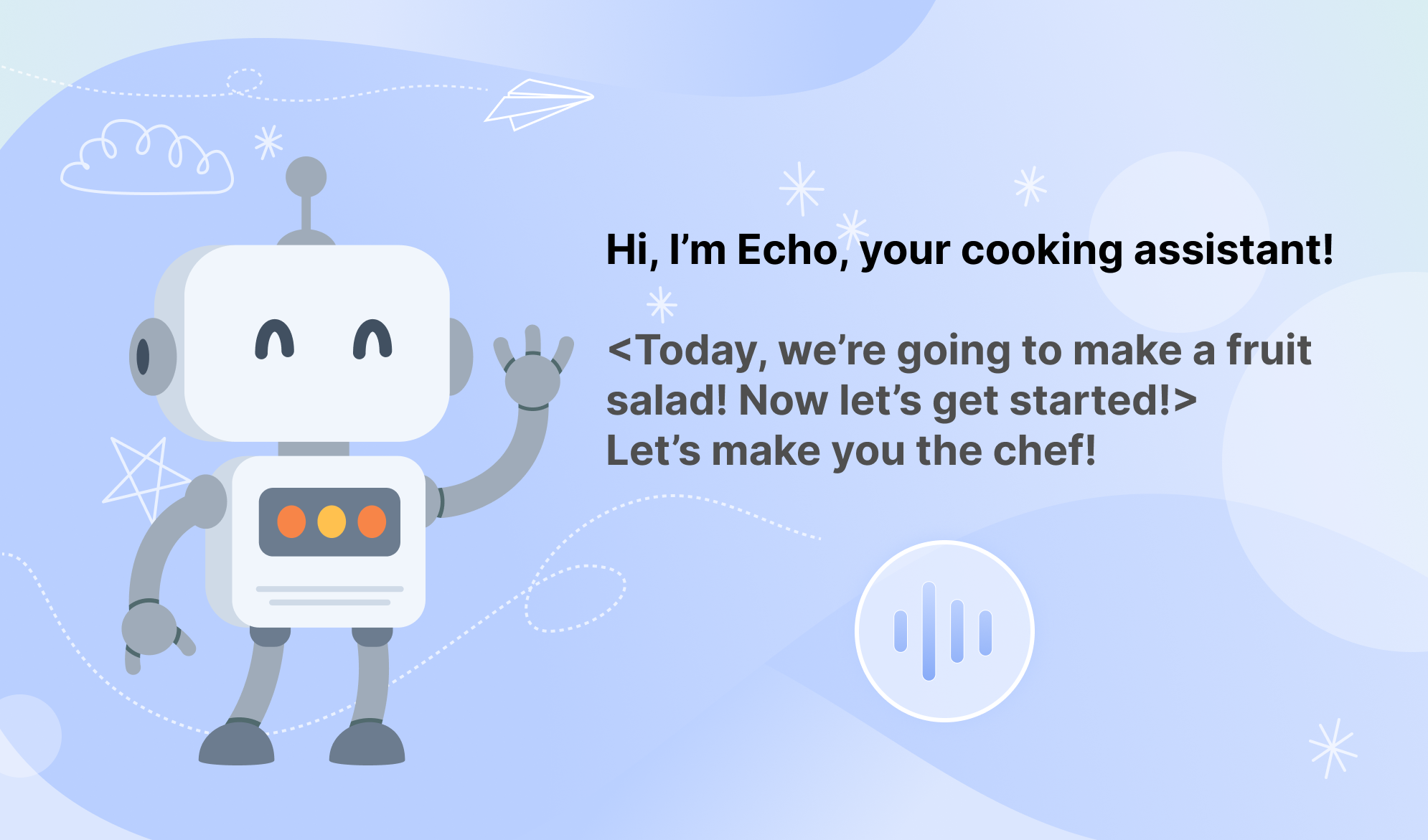}
    \caption{}
    \label{subfig:WoZ-ui}
  \end{subfigure}
\caption{Figure \ref{subfig:WoZ-setting}: Camera perspective in the WoZ experiment; the wizard observes the user's behavior through the camera. Figure \ref{subfig:WoZ-ui}: The UI of our WoZ system. The opening statement will appear on the screen at the beginning and disappear once the conversation starts.}
  \label{fig:WoZ}
\end{figure}

\subsubsection{Procedure}
We conducted a lab study with five new parent-child pairs using a kitchen toy set. Parents were asked to engage in open-ended, play-based English learning for 20 minutes without predefined goals.

The study used a Wizard-of-Oz setup in which a human ``wizard'' simulated an AI assistant. The wizard observed the family’s interaction and the parent’s screen use via a camera (Fig.~\ref{subfig:WoZ-setting}) and used a voice changer and a slight response delay to mimic an AI. The system was designed primarily for parent-facing use: parents were the main users who initiated interactions through a minimalist, single-button interface (Fig.~\ref{subfig:WoZ-ui}), while children did not independently operate the system. At the same time, because the wizard monitored the ongoing parent--child interaction, children’s speech and actions could still indirectly shape the wizard’s responses as part of the shared play context.

For preparation, the wizard drew primarily on materials recommended by T1. These included several reference EFL videos on key considerations in preschool bilingual teaching, as well as a classroom outline containing vocabulary lists and a branching cooking-game scenario. We adapted these materials to fit a 20-minute session, based on parents’ reports in the interviews about the upper bound of a typical single play session at home. During the study, these materials served as the wizard’s main supplemental guidance. As a graduate student who had passed China’s College English Test Band 6 (CET-6), the wizard could also use AI tools to quickly look up additional information when needed during the session.

The system was entirely passive, only responding to parents' queries and never offering unsolicited advice. Following each play session, we conducted a 40-minute semi-structured interview with the parent.

Thematic analysis~\cite{clarke2017thematic,Braun01012006} was conducted to draw insights from the experiment. We recorded and transcribed all WoZ sessions and follow-up parent interviews. Guided by our research questions and insights from prior interviews, two researchers first familiarized themselves with the data and collaboratively built a codebook that combined deductive (question-driven) and inductive (data-driven) codes. They then independently coded all transcripts and iteratively refined the codebook through discussion, resolving disagreements by consensus. Finally, we grouped related codes into higher-level themes that characterized parents’ use of AI-informed strategies, emerging tensions in AI-supported situations, and evolving expectations for AI, which directly informed our four design challenges and subsequent design principles.

\subsubsection{Findings}
Based on the WoZ experiment, we identified several key patterns in how parents interacted with the AI system. 

\textbf{ Utilizing AI for Foundational Support: Providing Vocabulary and Game Guidance: }
Parents expressed appreciation for the AI system's two main features: vocabulary recommendation and game step guidance. Both functions helped reduce hesitation in their speech and improved the flow of the game. This was especially helpful for less experienced parents (e.g., P1), for whom the AI's real-time support was crucial in maintaining their teaching confidence. As P1 remarked, \textit{``Whether I don't know a word or the child asks a 'tricky' question, after asking the AI, I immediately know how to continue.''} These functions helped parents feel more in control, as they had reliable resources to lean on when encountering language or game-related challenges.

\textbf{Adopting AI's Approach: Imitation of Expression Use: }
Parents and children with varying language proficiency used the AI's responses in different ways, often adapting their teaching approaches to imitate the AI's language. For example, P3 admitted, \textit{``It's exhausting to maintain the game while figuring out how to teach, so I unconsciously imitate what the AI says. ''} Parents with higher language proficiency, such as P5, were able to focus on the AI's reasoning process, using its suggestions as models for their own teaching strategies. \textit{``I saw that it suggested selecting ingredients and then chopping them for the salad, so when making spaghetti later, we followed the same process. For me, it's easier; for the child, she's willing to repeat the task she likes. ''} In contrast, P4, who had limited English proficiency, felt awkward repeating full sentences and could only manage to read out individual words, which restricted her ability to engage naturally with the AI's output.

\textbf{Struggling for Balance: Oscillating Between Pedagogical and Playful Roles: }
While our initial interviews revealed parents' struggle to fuse play and learning, this challenge became even more pronounced in the triadic setting of our WoZ experiment. Parents like P4 relied heavily on the AI for translation, leading to a more structured, forced learning experience. \textit{``My English level is poor, but I still want to teach my child something, so I keep asking the AI, then explain it to my child.'' } This overreliance on AI for language translation resulted in an overly directed teaching process that hindered the natural flow of play. 

P2, who lacked confidence in her language skills but had more experience with AI tools, faced a similar dilemma. Initially, she relied on the AI to provide English content but found this inadequate. She then shifted to guiding the child to discuss other issues within the scenario with the AI. Both tendencies highlight the difficulty parents face in maintaining a balanced integration of pedagogy and play during triadic interactions with the AI.

\textbf{Seeking a Partner, Not a Director: Parents' Call for Proactive AI Mediation:}
Parents' difficulty in regaining balance indicated a deeper need.
P1 reflected on the common struggle: \textit{``Starting from fixed teaching goals makes the game feel a bit forced, but I don't know how to get out of it.''} This not only illustrates that some parents became stuck in a rigid teaching mode, but also highlights the fact that parents are unable to adjust the interaction on their own. P4 further explained the clear need: \textit{``I just can't come up with other questions---the AI should remind me to add more fun elements.''} For many parents, especially those less familiar with using AI tools, this made it hard to formulate timely and effective prompts to guide the activity forward. 

Additionally, parents were keen to retain control over the process, emphasizing that the AI should serve as a playmate rather than dominate the interaction. As P3 noted, \textit{``At this age, I don't want the AI to dominate the interaction with the child; I need to control the whole process.''} P5 echoed this sentiment, saying, \textit{``The AI should introduce new content like a playmate, not take over the game.''} 
\\
\subsection{Formative Study Summary}
Our two-phase formative study first revealed parents' problems in play-based learning, then refined and validated the initial suggestions from teachers, and identified the impact of introducing AI into an open-ended, play-based learning environment through a WoZ experiment. The problems faced by parents can be summarized as four key challenges:

\textbf{Challenge 1(C1):} 
Parents struggle to identify relevant content, such as key vocabulary or game steps, that aligns with the child's developmental level and the immediate play context. This creates a \textit{content gap}.

\textbf{Challenge 2(C2):} 
Even when parents know what to teach, they often face a barrier in how to express it effectively, particularly if their language proficiency is limited. This \textit{expression gap} leads to hesitant speech and disrupts the interactional flow.

\textbf{Challenge 3(C3):} 
Parents find it difficult to dynamically balance between playfulness and instructiveness. This reveals a \textit{strategic gap}.

\textbf{Challenge 4(C4):} 
Parents often lack the ability to formulate effective prompts when they run into implicit problems. They can recognize the problems but cannot initiate a solution, creating an \textit{problem-solving gap}.

These challenges take into account the children's behavior and fully reflect the difficulties faced by parents, covering the spectrum from ``what'' to ``how''. In the following section, we will synthesize the challenges and findings throughout the formative study to derive our final design principles.

\section{Design of PAPEL}

\subsection{From Challenge to Design Principles}
In the previous section, we observed parent--child interactions step by step and identified four key challenges in play-based learning. 
Building upon previous parent-centric theoretical work, 
we propose four corresponding design principles to address these challenges:

\textbf{DP1. Provide Timely, Contextual Teaching Content and Game Guidance.} AI systems should offer relevant, well-timed instructional content and game progression suggestions. This form of \textbf{content support (C1)} directly fills the parent's knowledge gap, relieving the cognitive burden of inventing a curriculum on the fly and embedding learning seamlessly into gameplay.

\textbf{DP2. Provide Expression Support Through Language-Level Adaptation.} AI systems should tailor their linguistic output to match the parent's proficiency level. By providing this tailored \textbf{expression support (C2)}, the AI enables parents and children to speak more confidently and fluently. This not only facilitates smoother communication but also makes it easier for parents and children to imitate and adopt the AI's phrasing, allowing them to engage more naturally while still meeting their pedagogical goals.

\textbf{DP3. Balance Educational and Playful Goals.} The AI system should optimize the experience by actively balancing educational value with playfulness. This \textbf{goal support (C3)} ensures the interaction remains both engaging and pedagogically meaningful. By modulating its interventions, the system can sustain the child's engagement and the parent's instructional intent, preventing the interaction from becoming imbalanced.

\textbf{DP4. Offer Semi-proactive Guidance.} Given the parent's limited problem-solving ability and desire to lead the game, the system should offer \textbf{interactional support (C4)} through semi-proactive guidance. When it detects hesitation or stagnation, the AI should provide additional lightweight suggestions beyond the basic response. This helps parent--child dyads regain momentum while feeling in control, ensuring support is available without undermining their central role.


\vspace{1em}
Together, these four design principles offer a comprehensive framework for supporting parents through multiple dimensions of co-play: helping them access appropriate content, express it effectively, maintain a productive learning experience, and receive timely interactional guidance. Although developed for English language learning through pretend play, these principles have broader relevance for AI-supported, parent-led educational activities.

\subsection{System Design}
\label{subsec:system_design}
Building on insights from preliminary research, we designed and developed PAPEL, a chatbot intended to support the creation of an English language environment during parent--child play. 
PAPEL consists of four core LLM-driven modules. Table~\ref{tab:DP and SF} summarizes how these four modules and their key features map onto the design principles. 


\begin{table}[htbp]
\centering
\caption{Mapping of Design Principles to System Modules and Features. }
\label{tab:DP and SF}
\renewcommand{\arraystretch}{1.5} 
\begin{tabularx}{\textwidth}{|>{\centering\arraybackslash}m{6.4cm}|X|}
\hline
\rowcolor{gray!20}
\textbf{Design Principles} & \multicolumn{1}{c|}{\textbf{System Modules and Features}}  \\
\hline
\multirow{3}{=}{DP1. Provide Contextual Content Support.} & 
\begin{itemize}[leftmargin=*]
    \item M1. Content Generation
    \begin{itemize}
    \item Vocabulary Recommendation
    \item Play Guidance
    \vspace{1em}
    \end{itemize}
\end{itemize} \\
\hline
\multirow{3}{=}{DP2. Provide Adaptive Expression Support.} & 
\begin{itemize}[leftmargin=*]
    \item M1. Language Adaptation
    \begin{itemize}
        \item M2. Language Proficiency Analysis
        \item Adaptive Response Style
        \vspace{1em}
    \end{itemize}
\end{itemize} \\
\hline
\multirow{3}{=}{DP3. Balance Educational and Playful Goals.} & 
\begin{itemize}[leftmargin=*]
    \item M3. Balance Assessment
    \begin{itemize}
        \item Playfulness Score
        \item Educational Score
        \vspace{1em}
    \end{itemize}
\end{itemize} \\
\hline
\multirow{3}{=}{DP4. Offer Semi-proactive Guidance.} & 
\begin{itemize}[leftmargin=*]
    \item M4. Extended Response
    \begin{itemize}
        \item Intervention Planning
        \item Output Synthesis
        \vspace{1em}
    \end{itemize}
\end{itemize} \\
\hline
\end{tabularx}
\end{table}

\begin{figure}[htbp]
  \centering
  \includegraphics[width=1\linewidth]{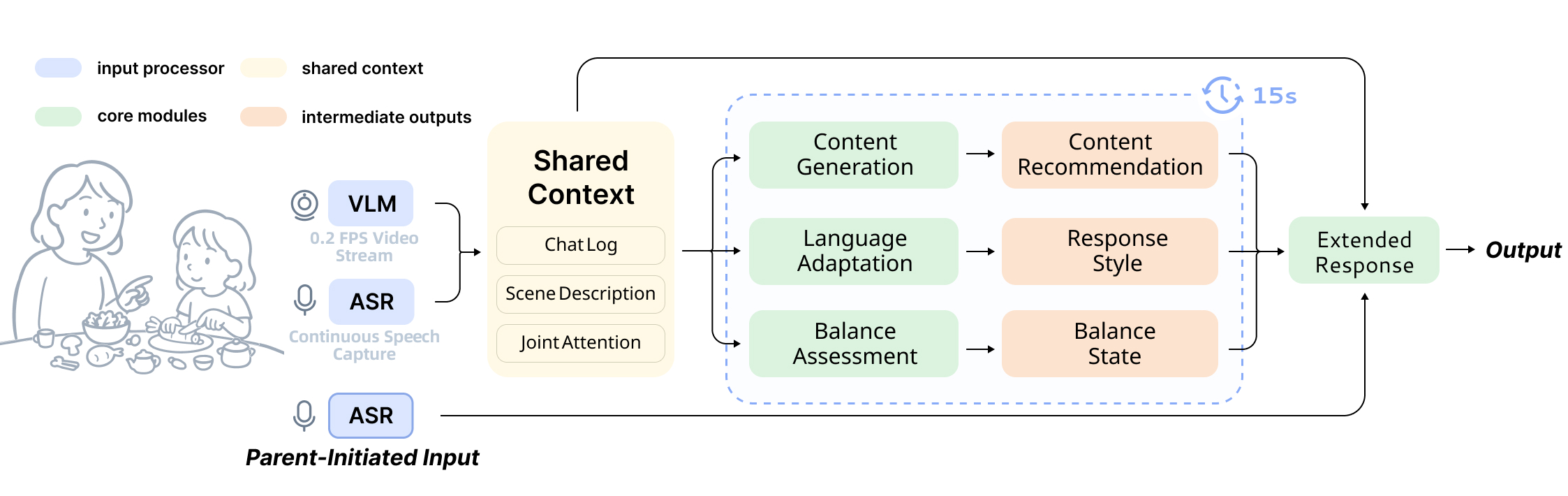}
  \caption{System overview of PAPEL. 
   Leveraging a VLM module and an ASR module, PAPEL first converts the background multimodal input into a semi-structured context that serves as the shared input to subsequent modules. Every 15 seconds, the Content Generation,
Language Adaptation, and Balance Assessment modules update the corresponding internal states.
The Extended Response Module then uses these intermediate outputs to produce system responses to parent-initiated input.
  }
  \label{fig:pipeline}
\end{figure}

As shown in Fig.~\ref{fig:pipeline}, PAPEL first converts the multimodal input into a semi-structured context that serves as the shared input to subsequent modules. The audio is transcribed with a commercial ASR tool. The video input is processed by a VLM at 0.2 FPS. Specifically, the VLM is prompted to generate a scene description for each frame. In addition to high-level descriptions, we also prompt the VLM to infer and analyze joint attention, which is key to understanding language acquisition in parent--child play~\cite{kwon2022captivate}. Here, we adapt the prompt template proposed by~\citet{shi2025towards}, which provides joint attention target and state descriptions of usable quality. Every 15 seconds, the Content Generation, Language Adaptation, and Balance Assessment modules update the corresponding internal states. The Extended Response Module then uses these updated results to produce system responses to parental input.

The \textbf{Content Generation Module} provides contextually relevant content by utilizing two LLMs to perform vocabulary recommendation and play guidance in parallel. The first LLM recommends five relevant preschooler-level words based on the shared context. The second LLM generates five play-oriented suggestions, either narrative or interactive prompts, by selecting from a set of prompt templates distilled from teacher and parent interviews, and then contextualizing the selected templates to the current play situation.
Notably, the prompt for these \textit{content recommendations} is designed to be \textbf{scenario-agnostic}, with only a single variable specifying the play theme, ensuring generalizability while allowing the generated content to remain focused on the current theme in practical use. Further details can be found in Appendix~\ref{appendix:implementation-content-gen}.

The \textbf{Language Adaptation Module} performs lightweight language proficiency analysis of recent dialogue, using code-switching patterns \cite{gardner2009code,gleason1973code} to distinguish word-level mixing, sentence-level mixing, and full English conversation. Based on these three levels, we derive corresponding \textit{response styles} with varying proportions of English, sentence lengths and complexity levels, and reliance on L1 scaffolding, which are used to guide the expression style of the extended response module. See Appendix~\ref{appendix:implementation-language} for response style details.

The \textbf{Balance Assessment Module} tracks how the interaction balances playfulness and knowledge-oriented instruction. The system computes two scalar scores in the range 0--2. The playfulness score $s_p$ summarizes two binary LLM judgments aligned with our two guidance types: whether the narrative is progressing smoothly ($s_n\in{0,1}$) and whether the conversation is sufficiently varied ($s_d\in{0,1}$); we set $s_p=s_n+s_d$. The educational score $s_e$ captures the intensity of the parent’s instructional language over a recent three-minute window, with $s_e=0$ for no instruction, $s_e=1$ for instruction that is meaningfully interwoven with the ongoing game, and $s_e=2$ for frequent or repetitive instructional moves. The module outputs a binary \textit{balance state} and flags a potential imbalance whenever $|s_e - s_p| \geq 1$, so that the system can adapt subsequent guidance by modestly strengthening the less-emphasized dimension. See Appendix~\ref{appendix:implementation-balance} for details on how $s_p$ and $s_e$ are defined and calculated.

The \textbf{Extended Response Module} performs two key functions through Chain-of-Thought prompting: intervention planning and output synthesis. For intervention planning, it infers parental intent (e.g., instruct, manage play), compares this with the system’s current balance state, and decides whether and how to intervene. For example, PAPEL optionally introduces playful elements to nudge the conversation toward equilibrium when overly instructional language is detected. For output synthesis, PAPEL integrates the shared context, content recommendations, the selected response style, and the intervention decision to produce the final response. More details are in Appendix~\ref{appendix:implementation-response}.

\begin{figure}[htp]
    \centering
    \includegraphics[width=1\linewidth]{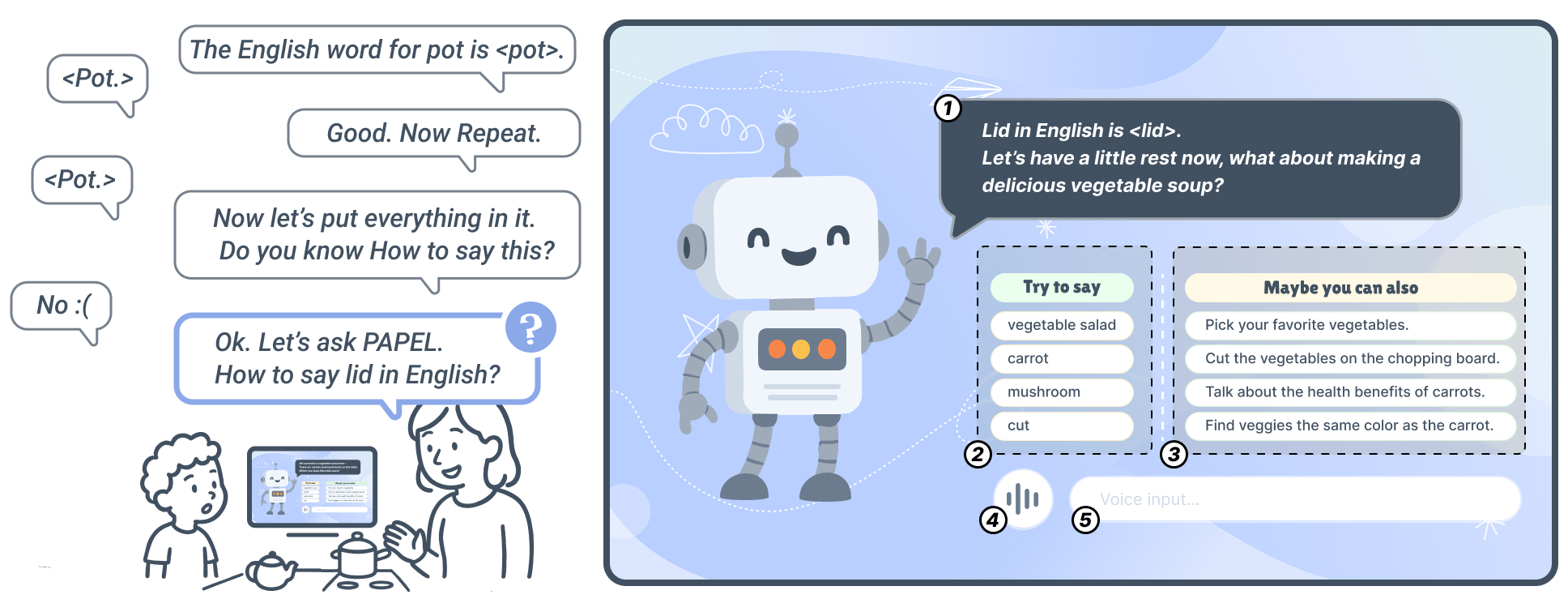}
    \caption{User interface of PAPEL. Key components include: (1) a response area at the top displaying PAPEL’s suggestions, (2) vocabulary cards that play pronunciation when tapped, (3) a panel showing rotating play guidance suggestions, (4) an interaction button that triggers system replies, and (5) a speech input box displaying real-time ASR transcripts of parent–child speech. Example conversation and PAPEL outputs were translated into English, and segments that were originally spoken in English are marked with angle brackets. See Appendix~\ref{appendix:conversation} for a real conversation transcript from the user study.}
    \label{fig:UI}
\end{figure}

Finally, the user interface (Fig.~\ref{fig:UI}) was deliberately designed in response to parents' concern about preschoolers' screen exposure, with the goal of reducing screen dependence and cognitive load. It includes the system response area, vocabulary cards, play suggestions, and a real-time transcription box with a voice input button. This setup maintains a child-friendly look while addressing parents’ concerns about excessive screen exposure, ensuring that attention remains on face-to-face interaction with the child. Due to space limitations, we do not provide detailed implementation methods for each component in the main text. Further details about PAPEL's system can be found in Appendix~\ref{appendix:implementation}.

\subsection{Function Validation and Optimization}

Before the main user study, following and extending prior work’s pilot-testing practice~\cite{wang2025lata,lee2022collabally}, we ran a pilot function validation with one volunteer parent–child dyad.
From the resulting logs, we obtained 13 parent queries and 53 module updates, on which we conducted two analyses: one on language proficiency labels and one on balance assessment with extended responses. For proficiency analysis, an English teacher who had participated in our interviews inspected each update and judged that the three-level proficiency labels were appropriate in all cases, indicating that GPT-4o could reliably support this classification task. For balance and extended response, two researchers independently analyzed each parent query and the corresponding system reply, then reconciled their judgments and member-checked them with the parent’s subjective experience in a follow-up interview. Across the 13 queries, they identified one clear instance of over-intervention and one case of insufficient intervention that the parent resolved by asking additional questions. Based on these findings, we iteratively refined prompts to further improve system performance before running the full study.

\section{User Study}
We conducted a within-subjects study with N = 16 participant dyads to evaluate our system. Our study focused on three main research questions:

\textbf{RQ1.} To what extent can PAPEL enhance the play-based learning process? 

\textbf{RQ2.} How do parents perceive the functionalities and features of PAPEL, and what are their overall experiences with the system?  

\textbf{RQ3.} In what ways, if any, does PAPEL support parents in facilitating their child's learning and engagement?  

Participants were asked to engage in two teaching activities within a game-based context, with each session using different toys and systems, for a total of approximately 2 hours. The experiment results were evaluated through session record analysis, subjective questionnaires, and semi-structured interviews. The following sections provide a detailed overview of the experimental design, participant recruitment, experimental procedures, and evaluation methods.

\subsection{Experiment Setup}

Our experiment adopted a within-subjects design with a baseline condition for comparison with PAPEL. Each dyad completed two sessions using different toy sets to reduce learning and fatigue effects on the child and limit carryover from the parent’s experience in the first session. To further control for these order effects, we followed the counterbalance scheme in prior work \cite{AutoPBL} and counterbalanced both the order of system exposure and the order of the toy set, yielding four experimental sequences. We therefore used quota-based recruitment to fill these four conditions evenly, resulting in 16 parent--child dyads in total (n = 4 per sequence).
\subsubsection{Game Scenarios}
We selected the two most frequently mentioned themes from the formative study—kitchen and clinic—as the basis for our experimental toy sets and purchased toy models that had been jointly recommended by the interviewed parents. 
As introduced in Section~\ref{subsec:system_design}, the themes were incorporated into the system prompts as contextual cues but were not accompanied by any additional content. 

\subsubsection{Baseline Setting}
To compare the effectiveness of our system on play-based learning outcomes, we established a baseline condition. We used Doubao \footnote{\url{https://www.doubao.com/chat/}}, a widely popular AI application among Chinese families \cite{Yu2025}, as the baseline system. Doubao demonstrates performance comparable to GPT-4o in Chinese \cite{superclue2025}, thereby making it an appropriate choice for our task.

In the baseline condition, we prompted Doubao to behave as a generic English-learning assistant for parent–child preschooler-level home-based English learning. To ensure comparability, we instructed Doubao to follow the same four design principles (DP1–DP4) that organize PAPEL’s functions (Table~\ref{tab:DP and SF}). We used Doubao’s official web interface as the baseline UI, which is nearly identical to a standard web-based LLM chatbot (e.g., ChatGPT Web): parents send messages via text or one-shot speech-to-text and receive turn-by-turn text responses. However, the web app does not support continuous speech or camera video streams, so the baseline system had no multimodal shared context of the ongoing play. This baseline approximates a typical form of LLM use that parents could plausibly access and configure at home at the time of our study.

\begin{figure}[ht]
  \centering
  \begin{subfigure}{0.3\textwidth}
    \includegraphics[width=\linewidth]{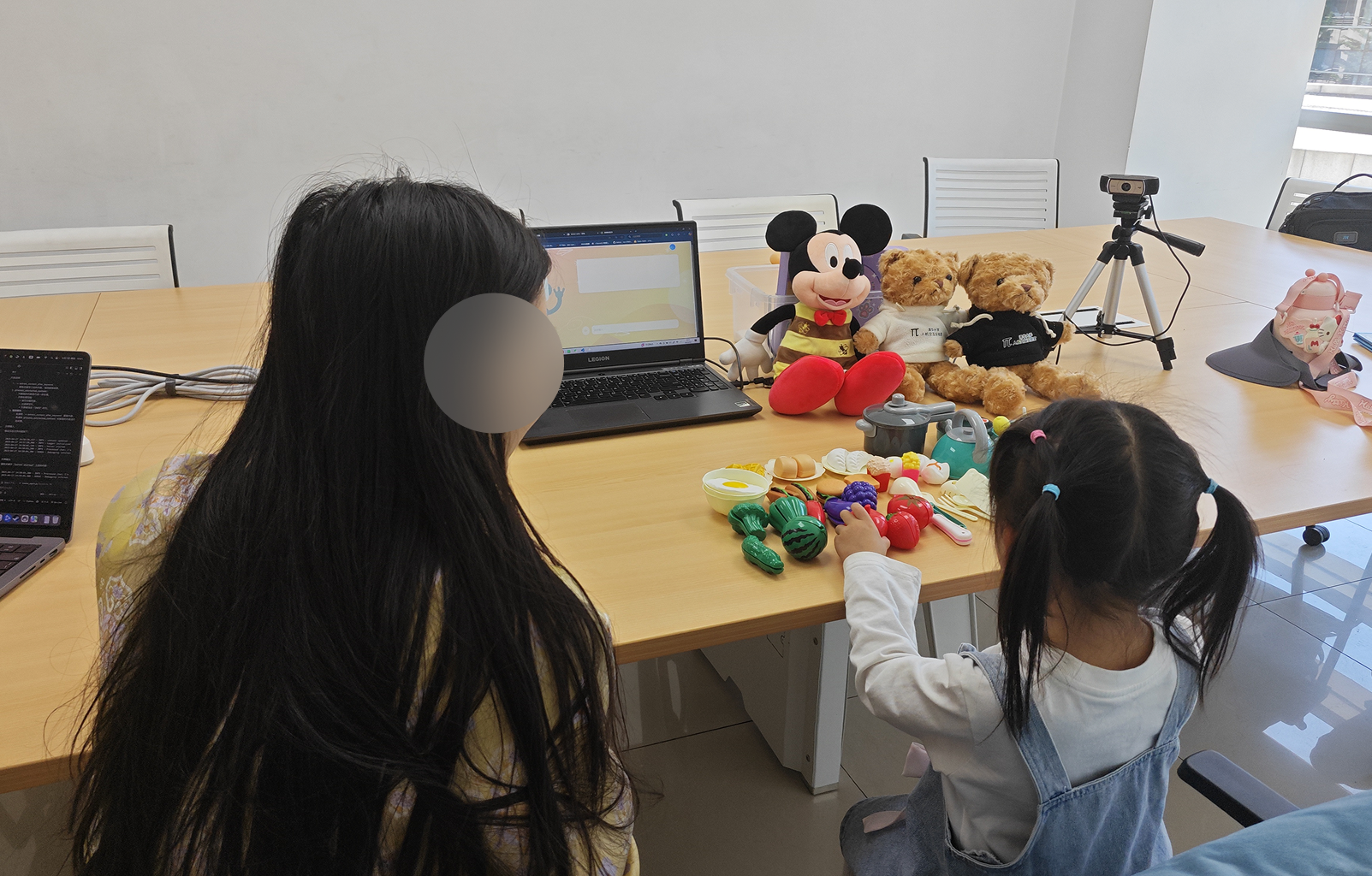}
    \caption{Experiment setup}
    \label{fig:Experiment setup}
  \end{subfigure}
  \hfill
  \begin{subfigure}{0.3\textwidth}
    \includegraphics[width=\linewidth]{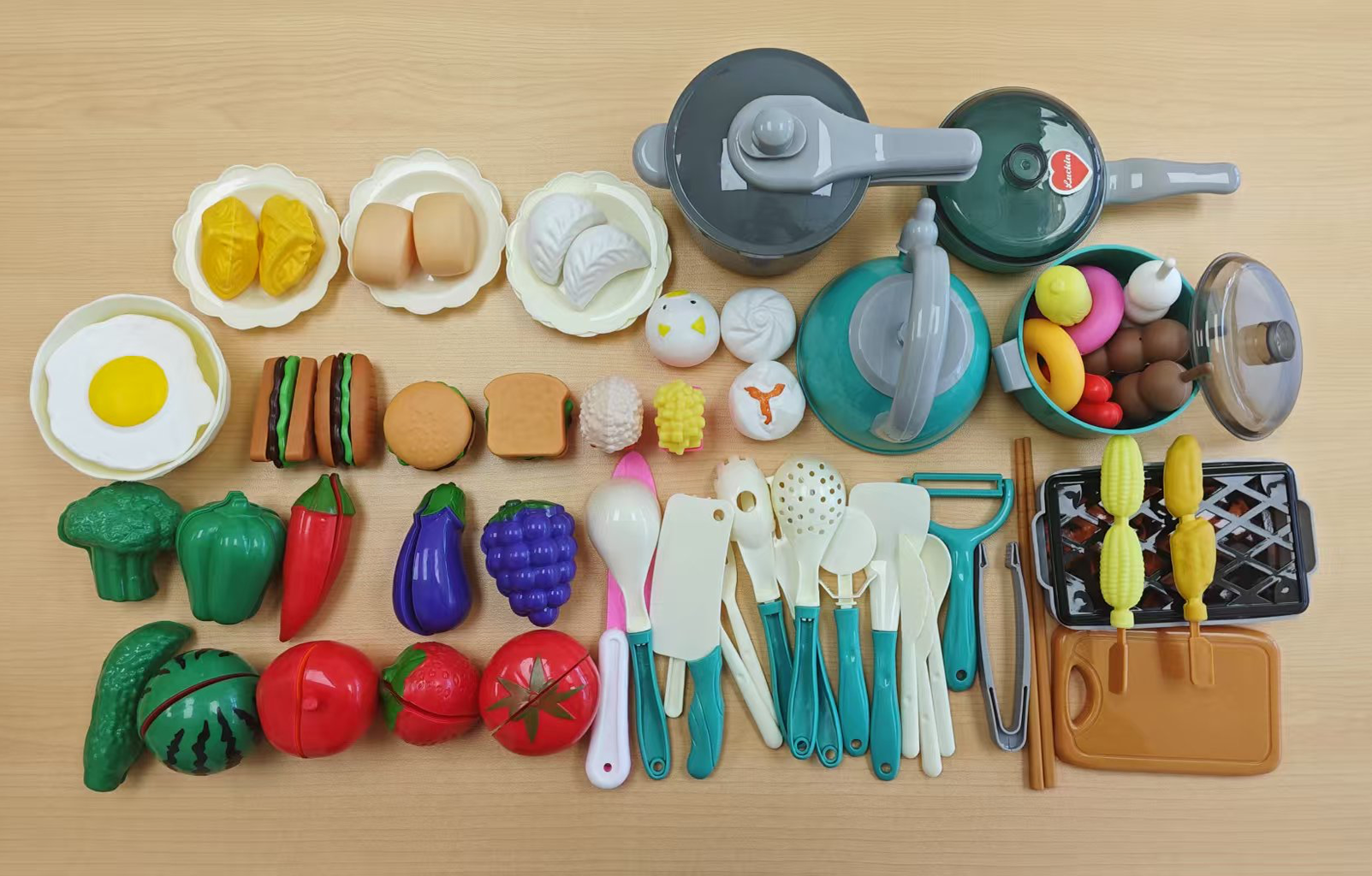}
    \caption{Kitchen toys}
    \label{fig:kitchen toys}
  \end{subfigure}
  \hfill
  \begin{subfigure}{0.3\textwidth}
    \includegraphics[width=\linewidth]{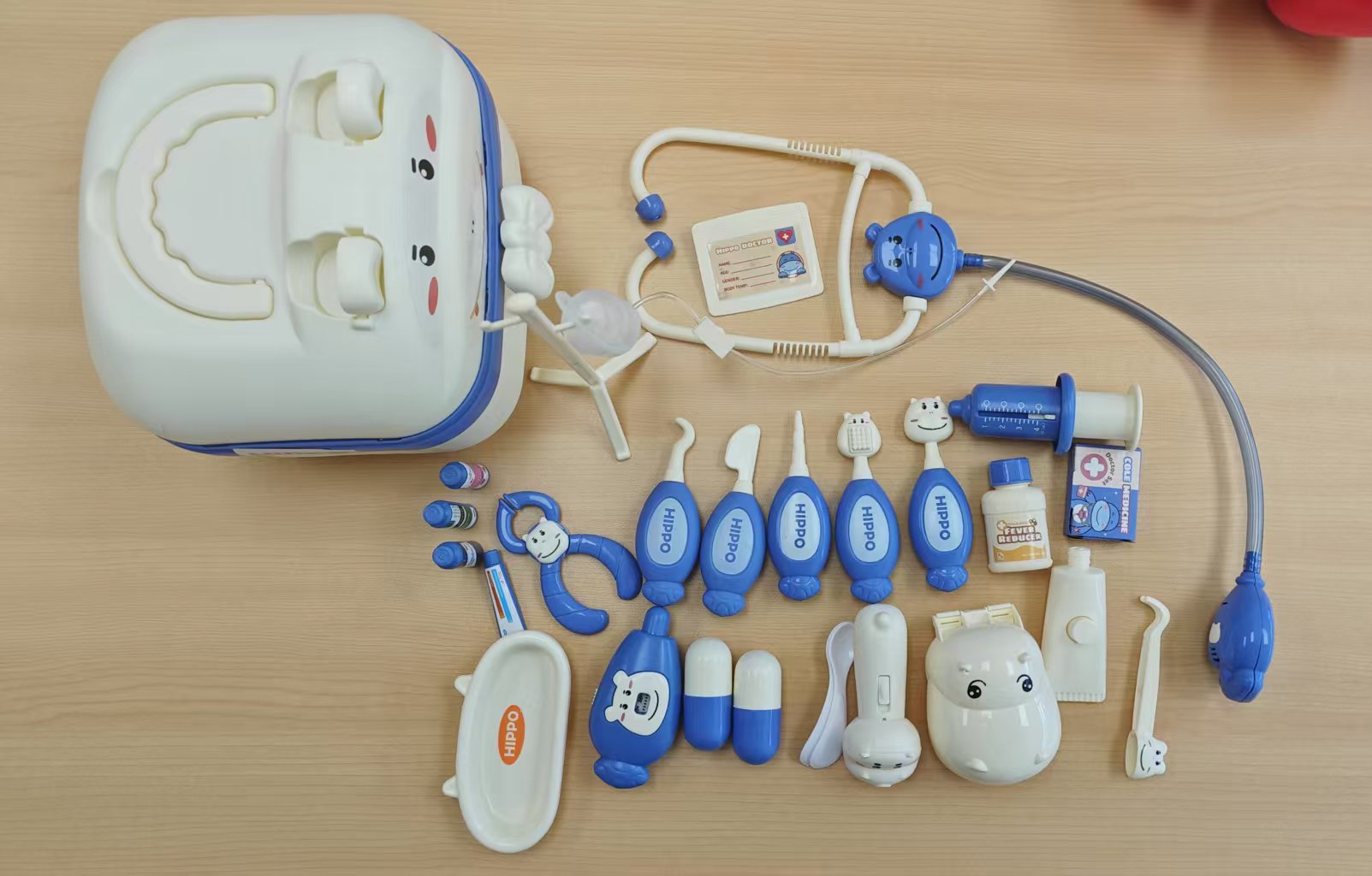}
    \caption{Doctor toys}
    \label{fig:Doctor toys}
  \end{subfigure}
  \caption{Experiment setup and toys used in the experiments.}
  \label{fig:three_subfigs}
\end{figure}

\subsection{Participants}
We recruited 16 participant dyads (Parents: 14F, 2M; Children: 11F, 5M) online. As shown in Table~\ref{tab:participant_info}, the children were aged 3--6 years old (mean = 4.6, SD = 1.1). There was no overlap in participants between the formative study and evaluation procedure. We did not set any specific language proficiency requirements for the parents and children, ensuring that we could capture the usage patterns of parents and children with varying language proficiency levels. In our study, we refer to the 16 participating parent–child dyads as D1--D16. The children and parents within each dyad are denoted as C1–C16 and P1–P16, respectively. To compensate for their time, each dyad received a \$20 incentive or an equivalent gift.

Our research was ethically reviewed by the Institutional Review Board (IRB). 
We informed both the parent and the child about the rules for data usage, privacy protection measures, and any potential risks on multiple occasions—first during the online contact phase and again before the formal in-person experiment began. They were explicitly told that participation was entirely voluntary and that they could withdraw from the study at any time if they encountered any difficulties. Parents were also informed that all video recordings would be used solely for research purposes, with identifying information removed, and that the recordings would be permanently deleted after the completion of the study. All participants who confirmed their involvement signed an informed consent form.

\begin{table}[ht]
\centering
\caption{Demographics of participants in the user study. Parent Eng. Prof. denotes self-reported English proficiency on a 1--5 scale. Child Eng. (mo) denotes the child's prior English-learning duration in formal classes in months. Parent educational background is reported as B.A. (Bachelor's), M.A. (Master's), and Ph.D. (Doctorate). Gender is coded as F (female) and M (male). Home Teaching indicates whether the parent reported prior experience conducting in-home English teaching activities with the child (Yes/No).}

\begin{tabular}{lccccccc}
\toprule
\textbf{ID} &
\makecell{\textbf{Parent}\\\textbf{Edu.}} &
\makecell{\textbf{Parent Eng.}\\\textbf{Prof.}} &
\textbf{Parent Gen.} &
\textbf{Child Age} &
\makecell{\textbf{Child Eng.}\\\textbf{mo}} &
\textbf{Child Gen.} &
\makecell{\textbf{Home}\\\textbf{Teach.}} \\
\midrule
D1  & M.A. & 3 & F & 6 & 32 & M & Yes \\
D2  & B.A. & 3 & F & 6 & 25 & F & Yes \\
D3  & M.A. & 4 & F & 5 & 1 & F & No  \\
D4  & M.A. & 2 & F & 5 & 3 & F & No  \\
D5  & B.A. & 3 & F & 6 & 6 & F & Yes \\
D6  & B.A. & 3 & F & 3 & 2 & F & No  \\
D7  & M.A. & 3 & F & 3 & 1 & F & No  \\
D8  & B.A. & 3 & F & 5 & 12 & F & Yes \\
D9  & Ph.D. & 5 & M & 4 & 10 & F & No  \\
D10 & B.A. & 2 & F & 6 & 12 & M & No  \\
D11 & M.A. & 3 & F & 4 & 1 & F & No  \\
D12 & M.A. & 4 & F & 3 & 3 & F & Yes \\
D13 & M.A. & 4 & M & 6 & 6 & M & Yes \\
D14 & B.A. & 1 & F & 4 & 3 & M & No  \\
D15 & M.A. & 2 & F & 4 & 6 & M & Yes \\
D16 & B.A. & 3 & F & 4 & 3 & F & Yes \\
\bottomrule
\end{tabular}
\label{tab:participant_info}
\end{table}

\subsection{Procedure}
Parent–child dyads participated in the study at a university-based research facility. To ensure consistent participation goals across families, we provided an overview of the study background and procedures. Drawing on guidance from the formative study (T2), parents were reminded to avoid turning the session into a rote vocabulary teaching task and were encouraged to incorporate language in a naturalistic and playful manner.

Each dyad completed two 20-minute play sessions in a room equipped with only the toys required for the current session and two sets of video recording equipment positioned to capture the interaction from multiple angles. Prior to each session, only the system and toy set for that session were introduced. To support parents in engaging meaningfully within the play scenario, the experimenter offered a brief orientation outlining possible directions the play could take, helping parents feel confident in initiating the activity.

Following each session, parents were asked to write down the content they intentionally delivered and complete a brief questionnaire evaluating their experience with the system. These responses provided immediate feedback on the interaction quality and informed subsequent qualitative analysis. 

The experimenter exited the room after the preparation to support naturalistic parent–child interaction. A 40-minute break was provided between the two sessions to reduce fatigue and allow for debriefing if needed. After completing both sessions, participants engaged in a semi-structured exit interview lasting approximately 30 minutes.

\subsection{Evaluation Methods}
\label{subsec:evaluation methods}
We evaluated the experiences of both parents and children through three complementary methods: analysis of session recordings, self-report questionnaires, and semi-structured interviews. In the following sections, we describe the evaluation metrics and data processing procedures associated with each of these methods. 
\subsubsection{Session Records Analysis and Learning Process Evaluation}
We collected and analyzed detailed records of each experiment session. All video recordings were first transcribed using automatic speech recognition, followed by manual corrections to ensure accuracy due to limitations in the recognition quality.

These recordings served as a primary data source for evaluating both instructional effectiveness and the quality of parent–child interaction during educational play. Guided by the recommendations from teacher interviews and in accordance with the literature \cite{pedagogical_play}, our research does not concentrate on the direct learning outcomes of the children. Instead, our focus is on whether PAPEL creates a more effective environment for learning through play. Therefore, rather than quantifying the vocabulary children memorized, our analysis shifted to evaluating the development of parents' teaching abilities and the shaping of the overall language environment. Our attention was directed toward three indicators: (1) instructional content richness, (2) lexical diversity in English output, and (3) the extent to which utterances integrated gameplay with instructional intent.

To assess instructional content richness, we analyzed post-session reports in which parents documented the number of intentional teaching points they aimed to deliver. Lexical diversity was quantified by calculating the number of unique English word types used per unit time by both parents and children, serving as a proxy for vocabulary exposure and usage breadth.

To evaluate integration between play and teaching, we conducted an utterance-level analysis. A total of 2507 parental utterances were initially labeled using GPT-4o into four mutually exclusive categories: (a) game-oriented, (b) instruction-oriented, (c) integrated or mixed-content, and (d) non-task-relevant. To develop the prompting scheme, two researchers first independently coded 189 utterances from one transcript into these four categories and reconciled discrepancies through discussion. We then used the consolidated annotations to design the GPT-4o prompt, incorporating examples from that transcript to clearly distinguish the four categories. When we applied the prompt to the same transcript and to another transcript containing 114 parental utterances for validation, agreement between GPT-4o and the two human coders reached Fleiss’ $\kappa$ values of 0.80 and 0.73, respectively, indicating substantial agreement \cite{Landis1977TheMO}. Finally, we used the number of integrated utterances as an indicator of the degree of integration between play and teaching.

In parallel, we assessed interactional quality through system-level and dyadic-level engagement metrics. These included the number of conversational turns between the parent-child dyad and the system, as well as the number of interactional turns between the parent and child during the session, adapted from prior work on parent–child-AI interactions~\cite{Social_robots_as_conversational_catalysts}. We conducted Wilcoxon signed-rank tests to assess statistical significance across all indicators above. This non-parametric test was chosen because our experiment followed a within-subjects design and the collected data did not meet the assumptions of normality.

To supplement the direct observation of the children, we also analyzed the video recordings with a focus on children's reactions. From this, we extracted several typical interaction cases among the parent, child, and PAPEL to further analyze the effectiveness of PAPEL's support, which will be discussed in Section~\ref{subsec:interview}.

\subsubsection{Subjective Measures}
To capture participants' subjective experiences and perceptions, we administered a post-session questionnaire comprising three sections. The first section included three items adapted from the Technology Acceptance Model (TAM)\cite{TAM} to assess users' perceived usefulness and ease of use, reflecting their overall acceptance of the system.
The second section was adapted from the NASA Task Load Index (NASA-TLX) \cite{NASA-TLX} and the Brief Perceived Self-Efficacy Scale (BPSES) \cite{BPSES}. It focused on users' perceived effort, levels of frustration, and self-efficacy during the gameplay and instructional processes.
The third section consisted of four items, each corresponding to one of our core design principles. Participants were asked to evaluate whether the system effectively supported them in addressing the associated challenges. Wilcoxon tests were also conducted across all subjective measures.

\subsubsection{Semi-structured Interview and Thematic Analysis}
After both experimental conditions, we conducted 40-minute semi-structured interviews with each participating parent to gather in-depth feedback on their experiences and to supplement our primary research findings. 

Following the thematic analysis methods \cite{clarke2017thematic,Braun01012006}, our process began by collaboratively coding one parent's interview transcript to develop an initial codebook. The remaining transcripts were then coded independently by two researchers. Each coded transcript was subsequently reviewed by the other researcher, and any disagreements were discussed until a consensus was reached. Finally, we synthesized these codes through an iterative process of thematic analysis, manually clustering them to identify recurring and meaningful patterns, which yielded 224 codes. To preserve fidelity, all analysis was performed in the original interview language, with the findings translated into English for this paper using the GPT-4o model (privacy mode).

\section{Results}
We present the results of the previously introduced evaluation in three subsections. We first present quantitative evidence demonstrating the effectiveness of our system in supporting the play-based learning process (RQ1). We then draw on subjective scale data to describe parents' overall impressions of system use (RQ2). Finally, we conduct thematic analysis to illustrate the specific types of support our system provided to parents (RQ3). Quotes from child and parent participants are included in italicized text, identified as \textit{C\#} and \textit{P\#}, respectively. 

\subsection{Quantitative Analysis: Effectiveness in Supporting Play-based Learning}
\label{subsec:quantitative analysis}
\begin{figure}[H]
  \centering
  \begin{subfigure}[b]{0.19\textwidth}
    \includegraphics[width=\linewidth]{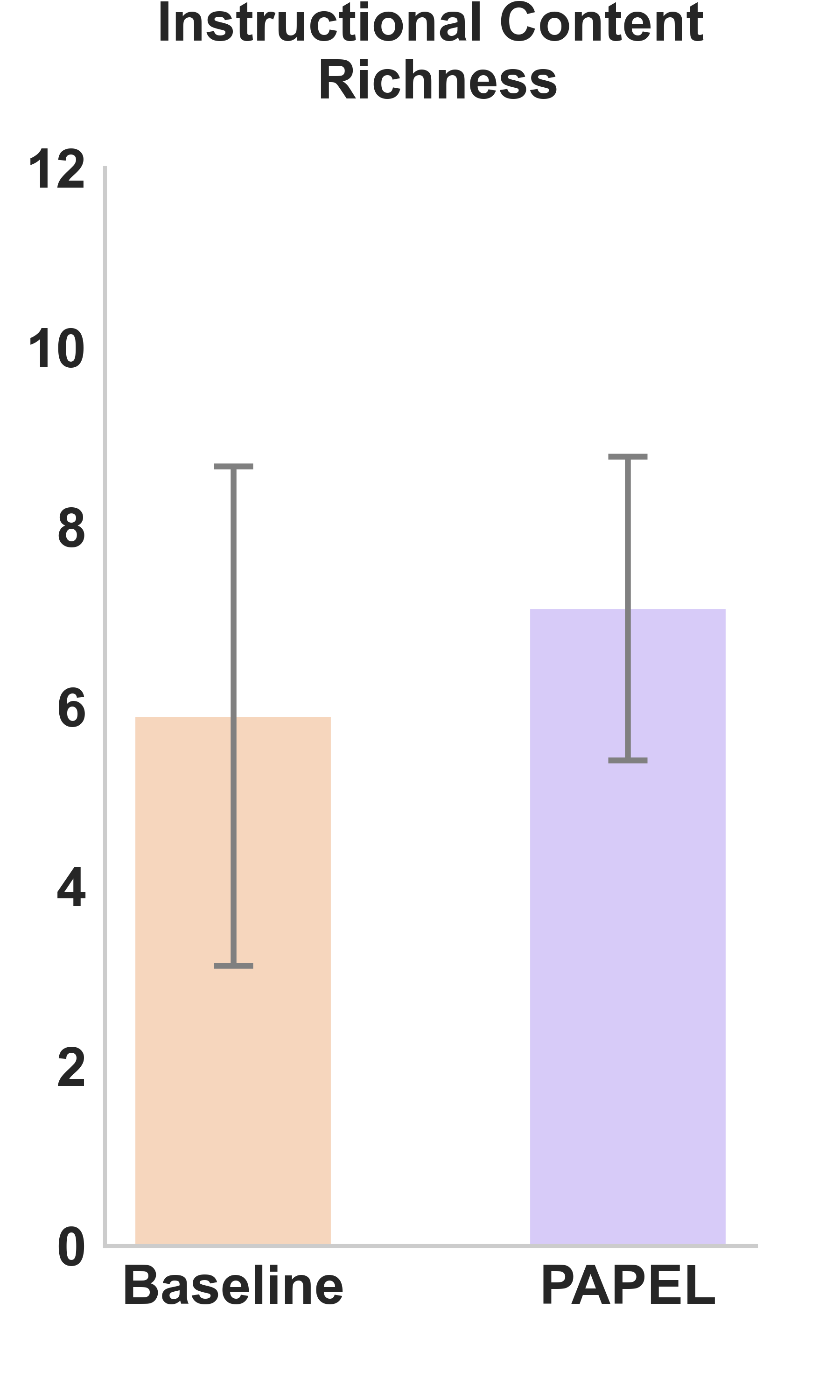}
    \caption{}
    \label{subfig:intentional teaching content}
  \end{subfigure}
  \begin{subfigure}[b]{0.19\textwidth}
    \includegraphics[width=\linewidth]{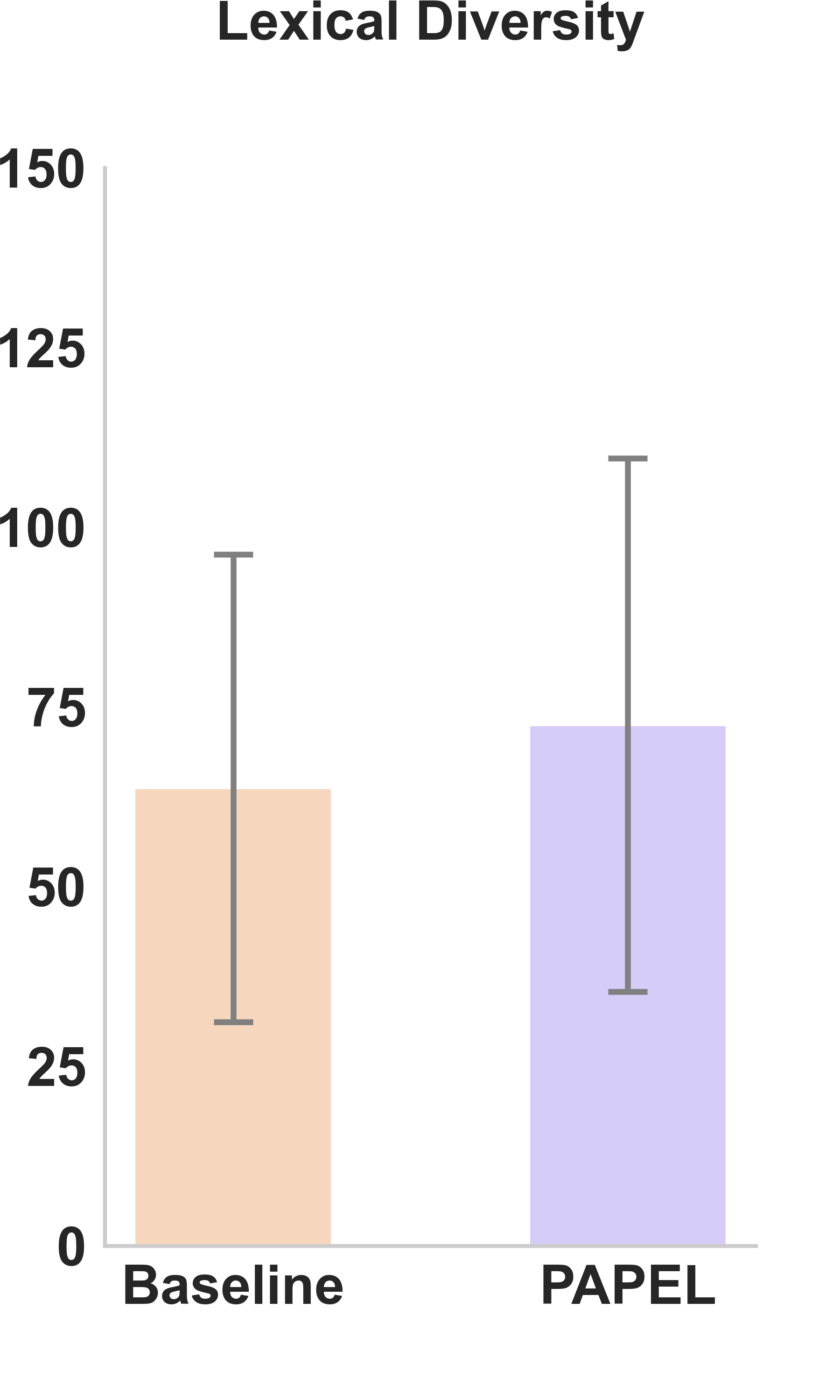}
    \caption{}
    \label{subfig:diversity}
  \end{subfigure}
  \begin{subfigure}[b]{0.19\textwidth}
    \includegraphics[width=\linewidth]{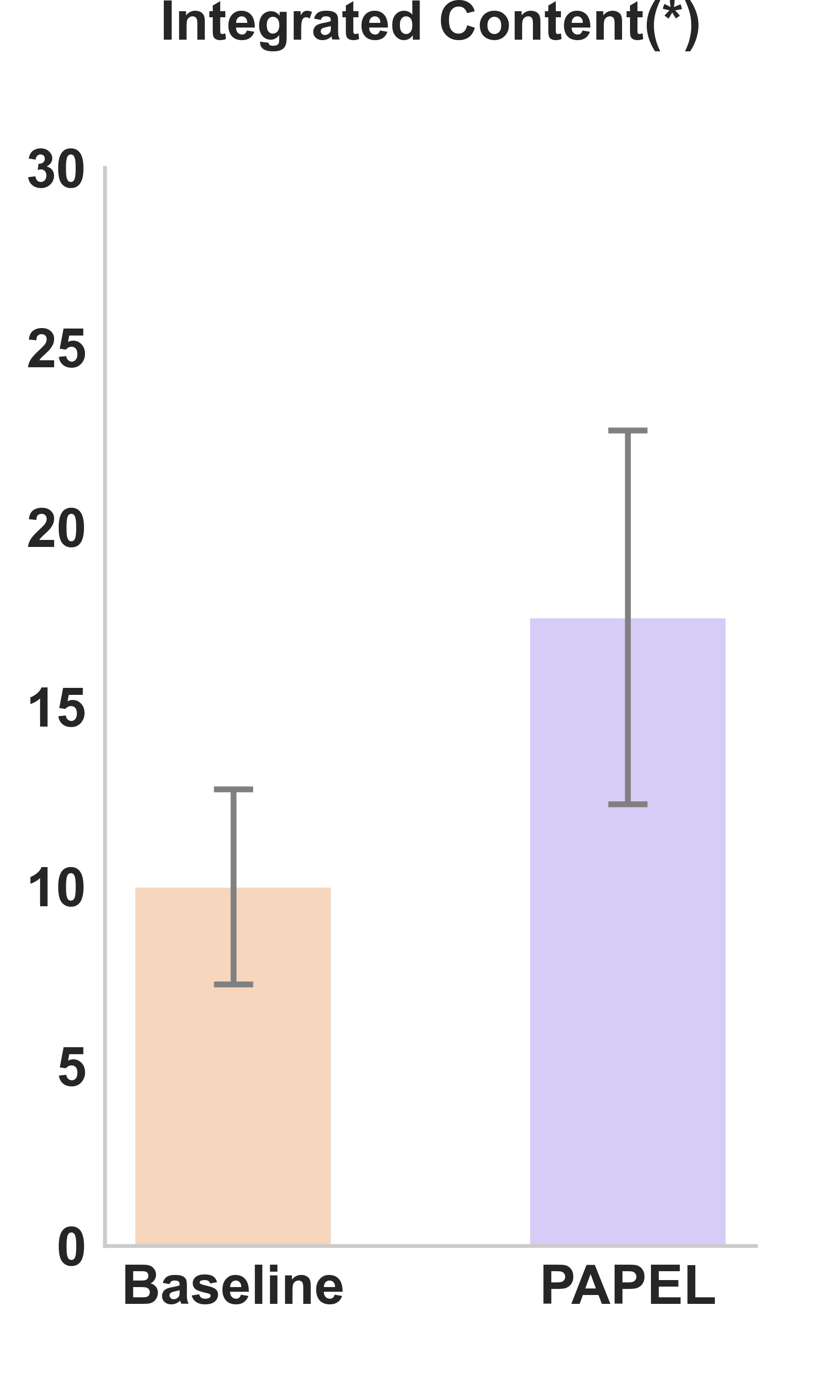}
    \caption{}
    \label{subfig:intergrated content}
  \end{subfigure}
  \begin{subfigure}[b]{0.19\textwidth}
    \includegraphics[width=\linewidth]{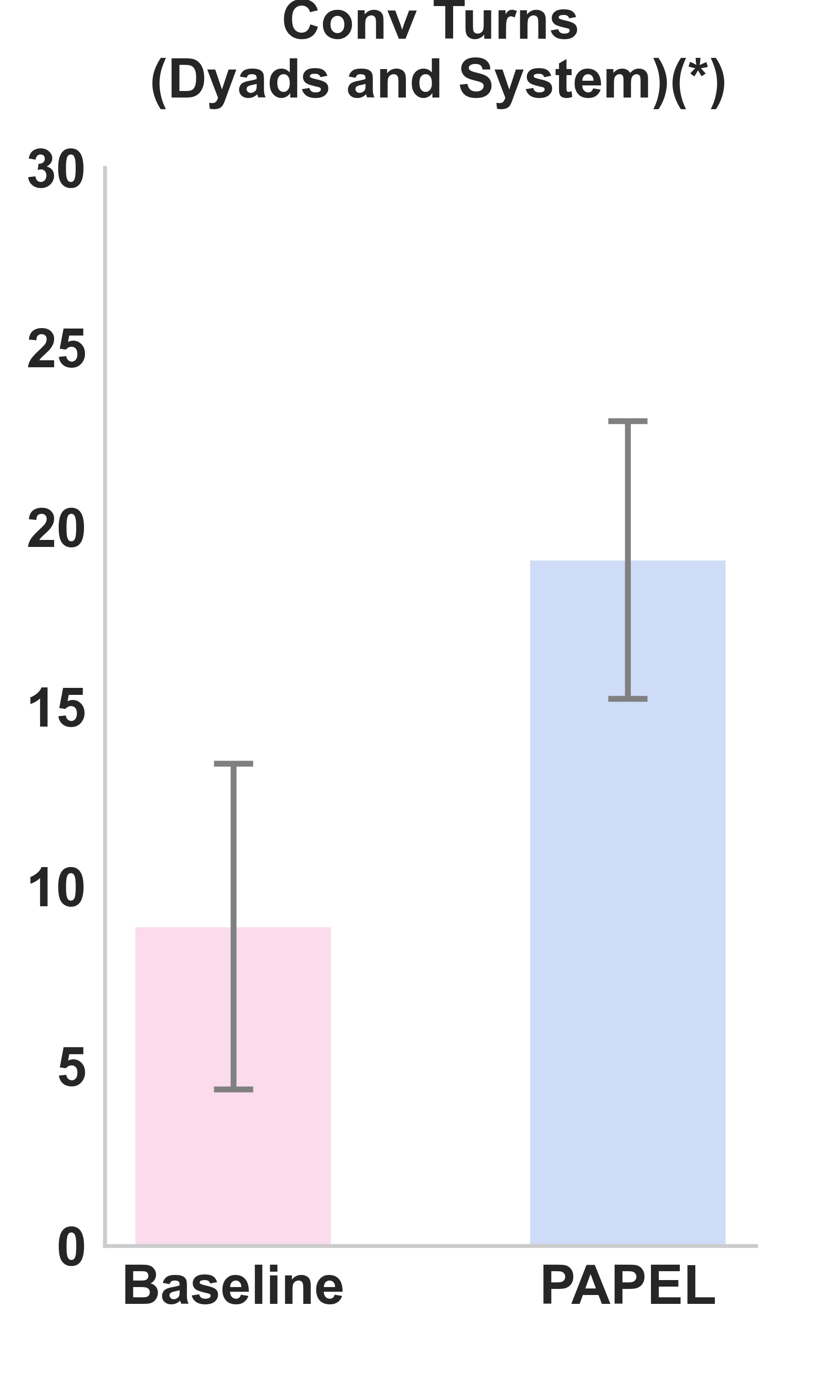}
    \caption{}
    \label{subfig:conv turns system}
  \end{subfigure}
  \begin{subfigure}[b]{0.19\textwidth}
    \includegraphics[width=\linewidth]{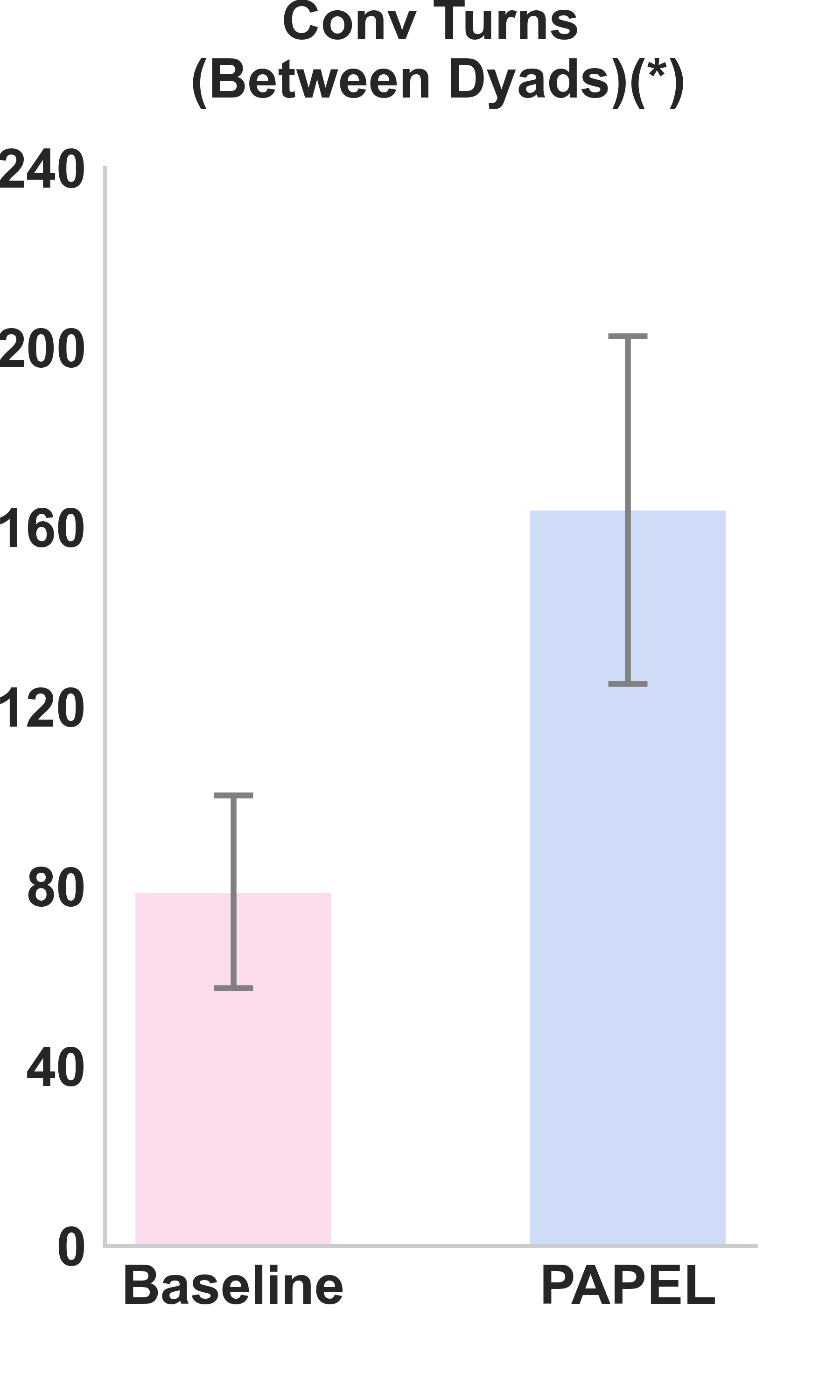}
    \caption{}
    \label{subfig:conv turns dyads}
  \end{subfigure}
  \caption{Session record analysis for instructional content richness, lexical diversity in English output, playful and instructional content integration, and conversation turns. These metrics reflect overall instructional effectiveness and interaction quality of sessions. Error bars indicate 95\% confidence intervals. Metrics with statistically significant differences are marked with an asterisk (*) in the figure title.
  }
  \label{fig:session-analysis}
\end{figure}

As shown in Fig.~\ref{subfig:intentional teaching content}, despite considerable variation in English proficiency among parent-child dyads, the number of intentional instructional points delivered by parents showed a slight difference when using our system (5.85 vs. 7.23), though statistical significance was not reached ($p=0.058$).
As P4 noted, ``\textit{The vocabulary recommendation feature helped me generate more divergent ideas---I didn't have to search for words myself or think of a word and then figure out a game to match it. It allowed me to quickly deliver richer teaching content.''}

Moreover, as shown in Fig.~\ref{subfig:conv turns system} and Fig.~\ref{subfig:conv turns dyads}, both parent-child and dyad-system conversational turn counts increased significantly when using PAPEL (parent-child: 78.9 vs. 163.9, p < 0.05; dyad-system: 8.9 vs. 19.2, p < 0.05). This suggests that the system benefited the flow and engagement of interaction, making the experience more dynamic and enjoyable for all participants. 

\begin{figure}[H]
  \centering
  \begin{subfigure}[b]{0.19\textwidth}
    \includegraphics[width=\linewidth]{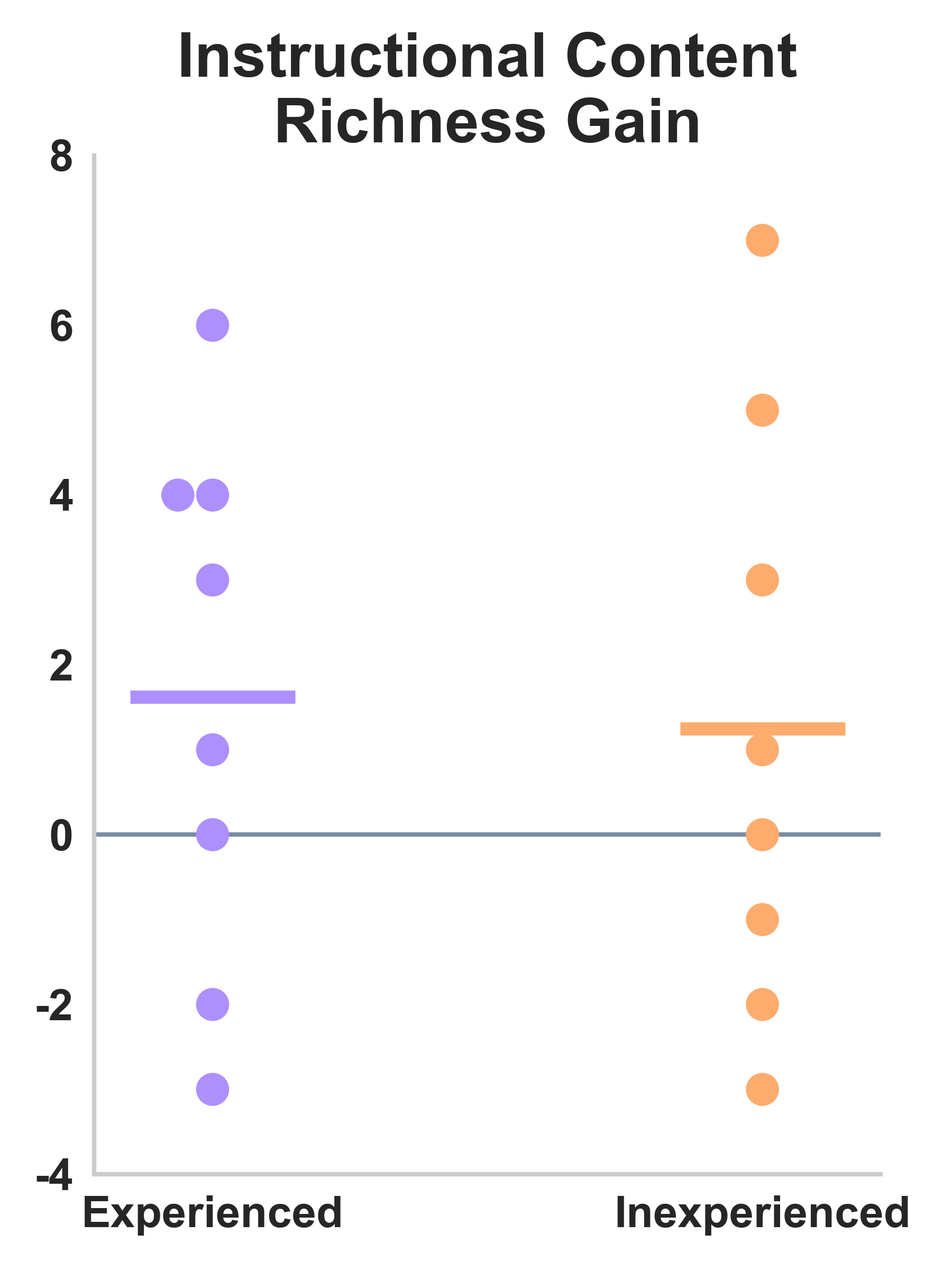}
    \caption{}
  \end{subfigure}
  \begin{subfigure}[b]{0.19\textwidth}
    \includegraphics[width=\linewidth]{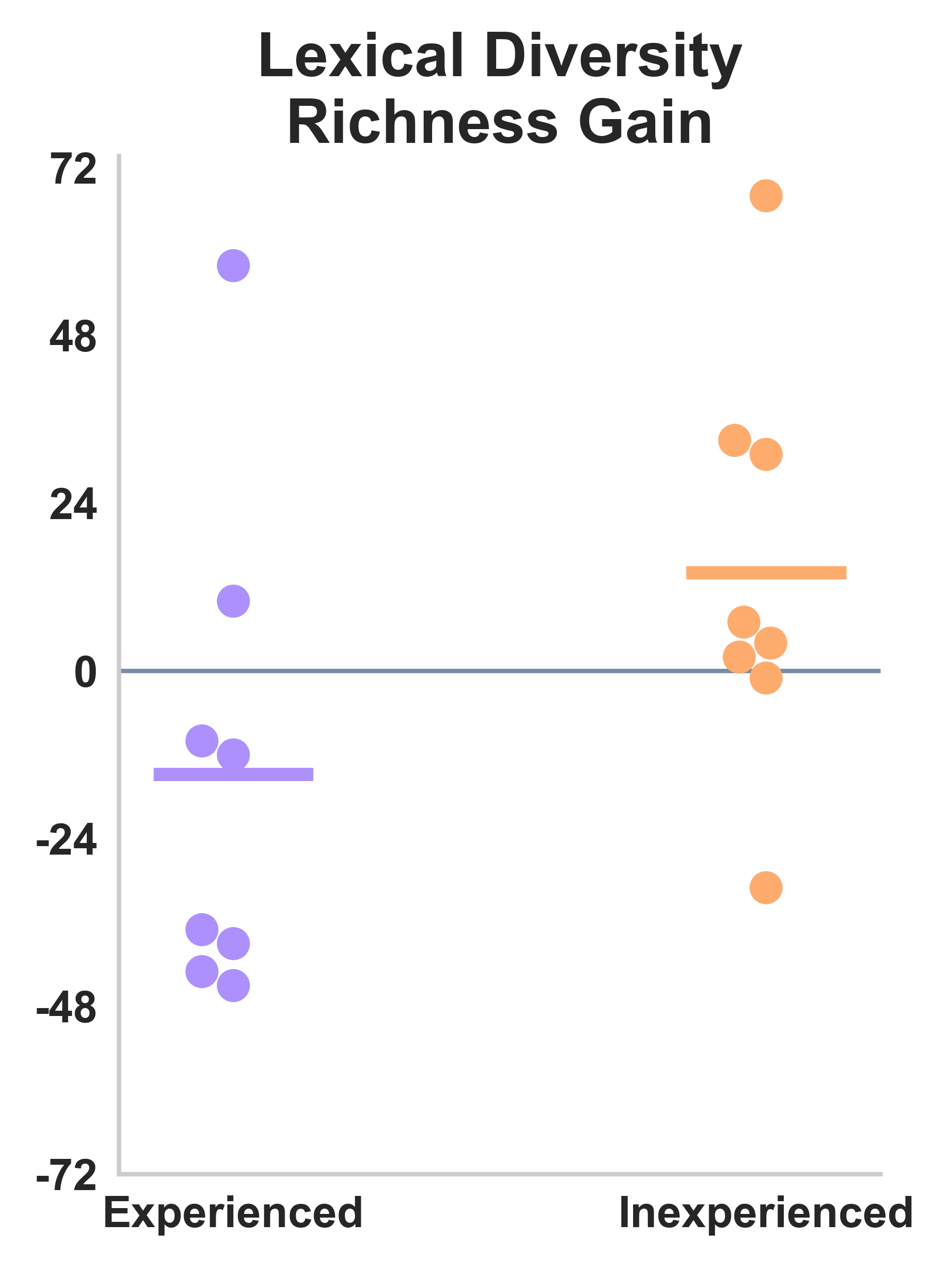}
    \caption{}
  \end{subfigure}
  \begin{subfigure}[b]{0.19\textwidth}
    \includegraphics[width=\linewidth]{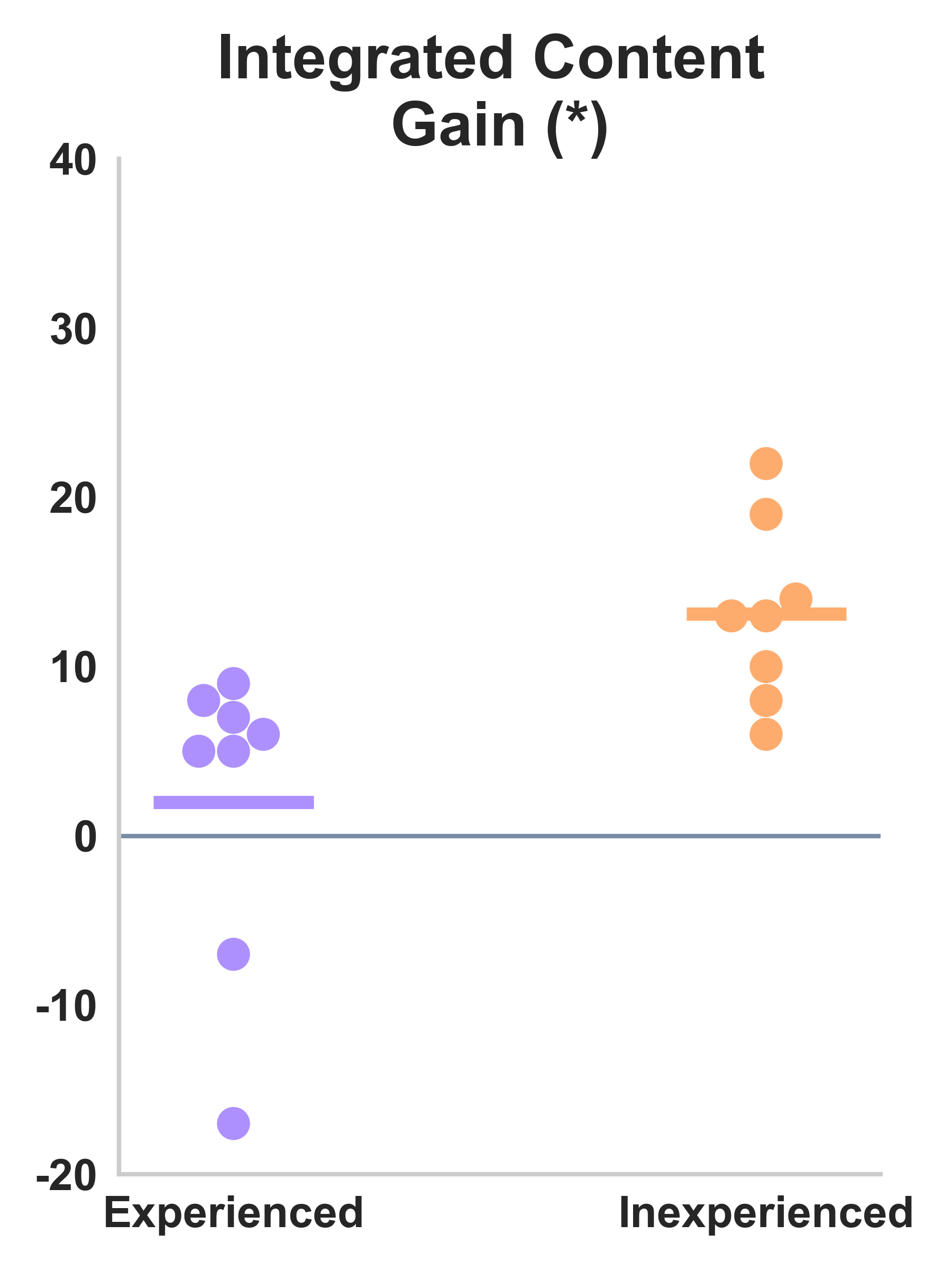}
    \caption{}
  \end{subfigure}
  \begin{subfigure}[b]{0.19\textwidth}
    \includegraphics[width=\linewidth]{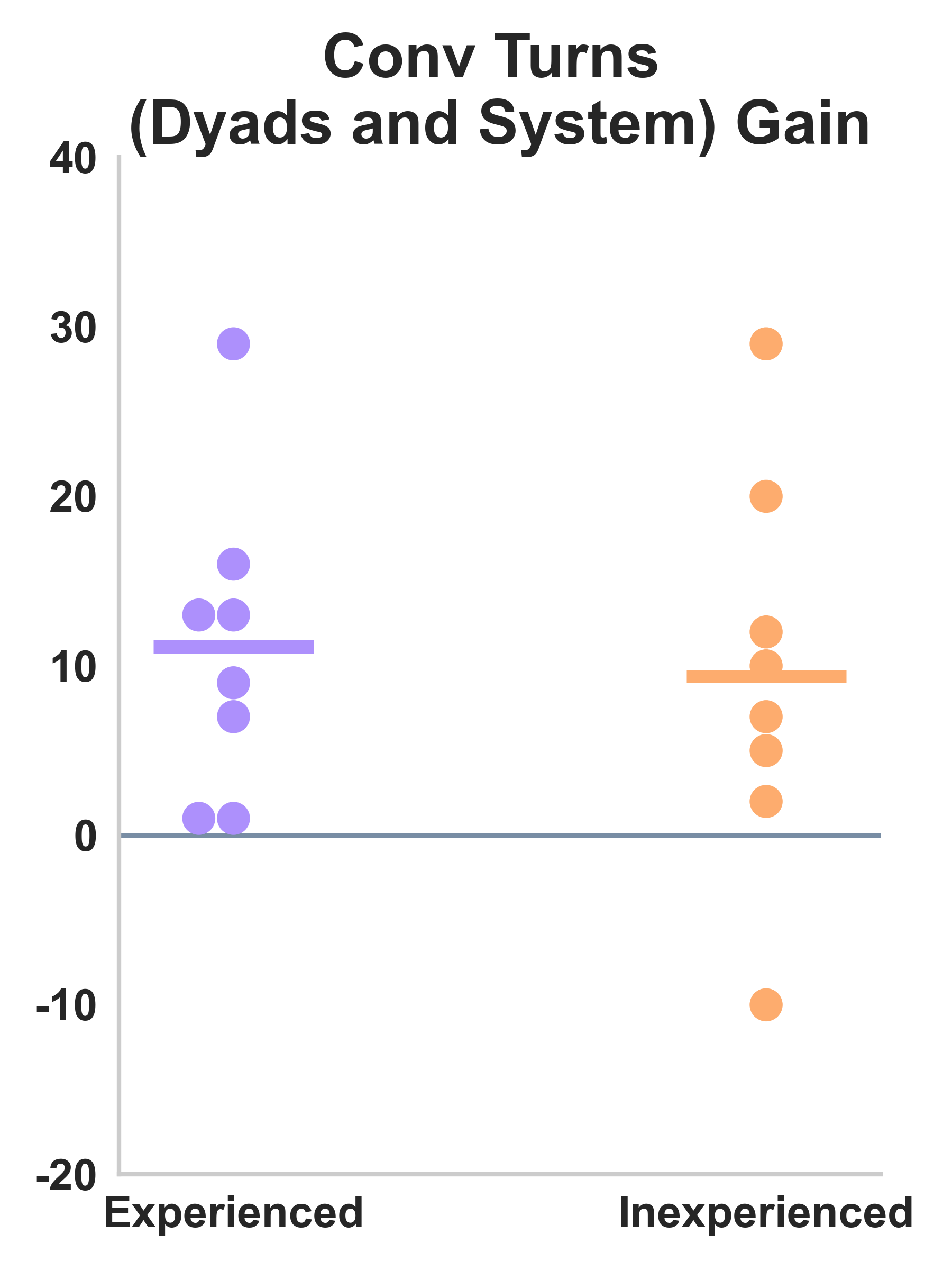}
    \caption{}
  \end{subfigure}
  \begin{subfigure}[b]{0.19\textwidth}
    \includegraphics[width=\linewidth]{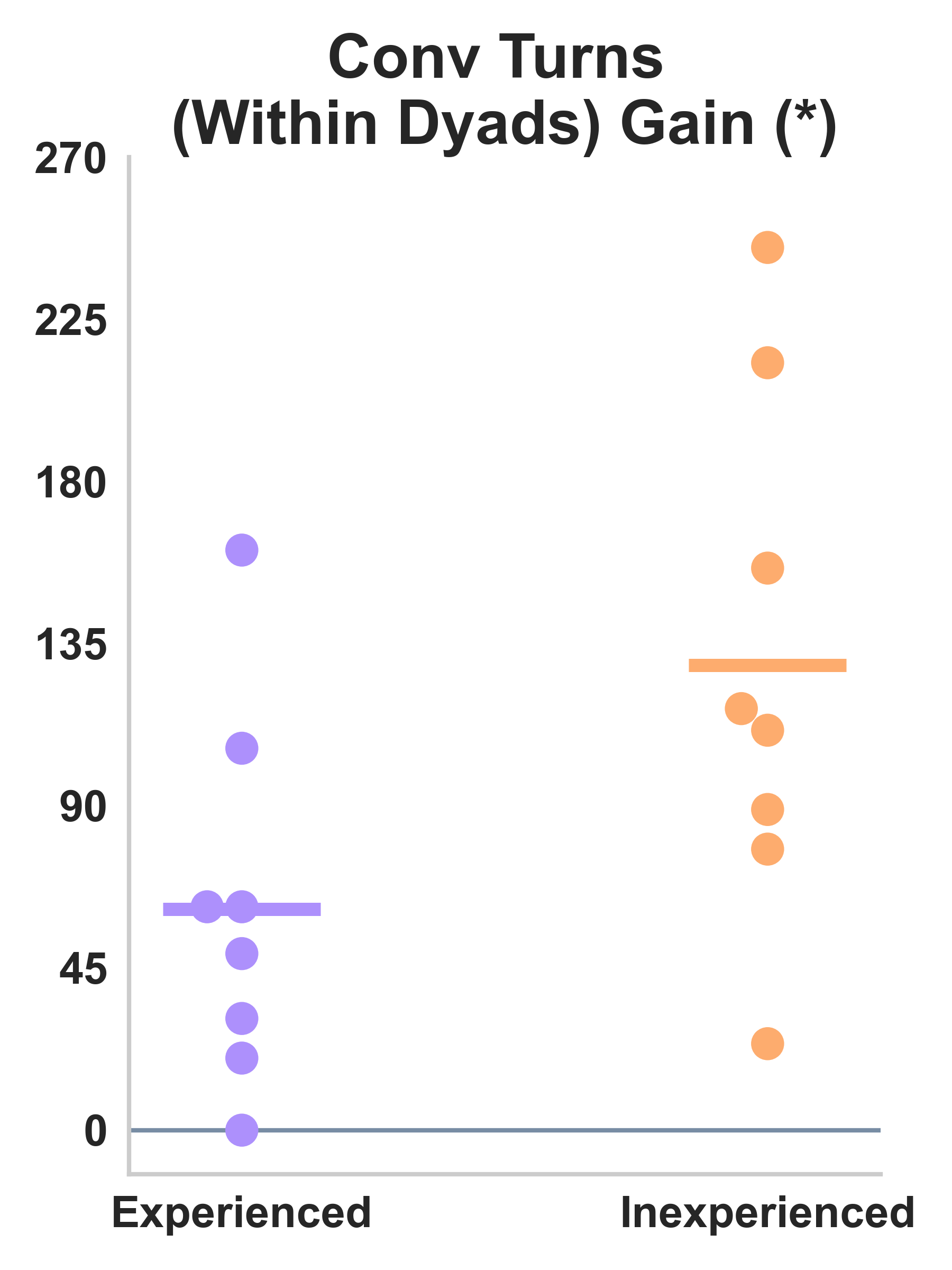}
    \caption{}
  \end{subfigure}
  \caption{PAPEL’s advantages over the baseline, comparing experienced and inexperienced parents. Inexperienced parents showed significantly larger gains in integrated content and conversation turns.}
  \label{fig:exp-analysis}
\end{figure}

When we further examined PAPEL’s effects by parents’ prior in-home learning experience, we observed several interesting patterns. As shown in Fig.~\ref{fig:exp-analysis}, PAPEL benefited parents overall, while parents without prior experience appeared to benefit more on certain measures, with significant differences particularly in integrated content and parent–child conversation turns (Mann–Whitney U test, p < 0.05).

Further analysis of their session recordings revealed differences in usage strategies. Across both conditions (PAPEL and baseline), parents with prior in-home teaching experience tended to treat AI as a content provider while retaining full control over the instructional flow themselves. In this context, parents assumed a traditional ``teacher'' role and replicated classroom-like interactions. For instance:
\\
\\
\begin{itshape}
\hspace*{1em}
P13: Repeat after me. What else did you do before stomachache?

C13: What else did you do before stomachache?

P13: Mind the details in pronunciation. What else did you do before the stomachache?

C13: What else did you do before stomachache?

P13: Good. \ldots
\end{itshape}
\\

In contrast, when interacting with PAPEL, most parents without prior in-home teaching experience adopted a more playful and relaxed role, positioning themselves more as co-players. Interestingly, P3 and P8---despite also having teaching experience---did not exhibit such divergence between conditions. In post-study interviews, both reported having attended parent training programs that emphasize using English naturally within play-based interactions, an approach conceptually aligned with our system design. As a result, their interaction style remained consistent across both systems. 

Together, these findings suggest that parents' self-perceived instructional role and their mental model of the teaching process significantly influence how they engage with AI-assisted systems, ultimately affecting learning outcomes. This indicates the need to support parents' reflection on their own roles. We will further discuss this phenomenon in the Discussion.

\subsection{Parental Perception on PAPEL}
\label{subsec:subjective feedback}
\begin{figure}[htbp]
    \centering
    \includegraphics[width=0.95\linewidth]{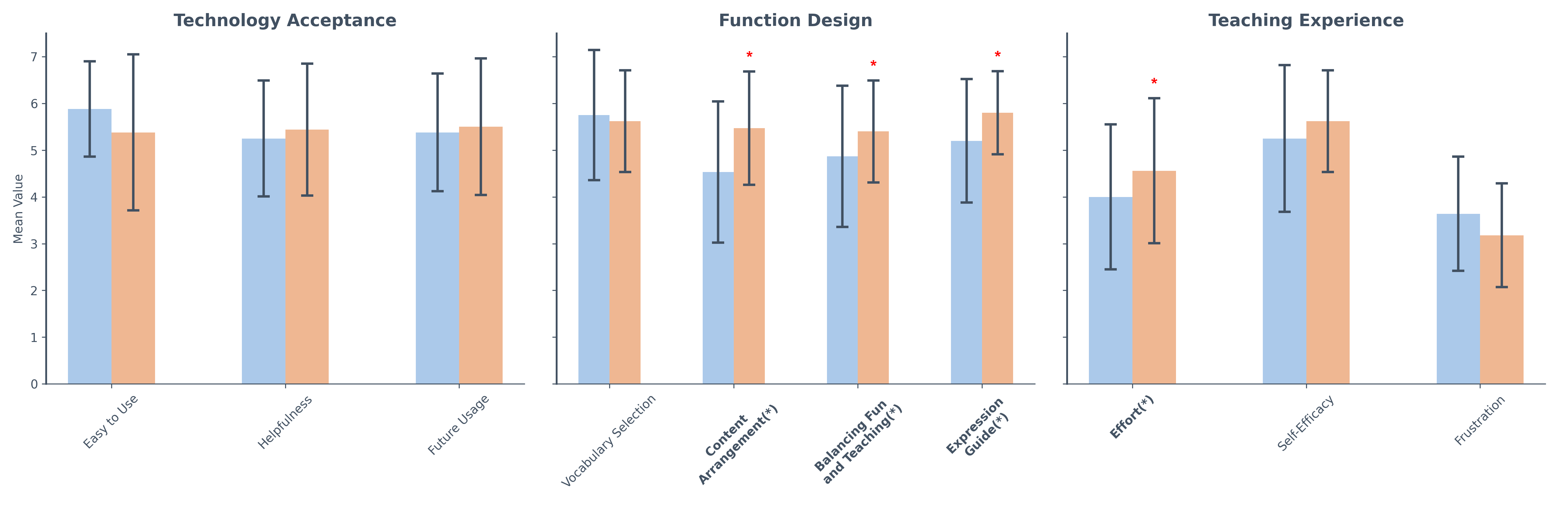}
    \caption{Subjective measures of parents for learning and play. Error bars indicate standard deviations. Parents perceived clear differences between PAPEL and the baseline in content arrangement, balancing fun and teaching, expression guidance, and effort. Metrics with statistically significant differences are marked with an asterisk (*).}
    \label{fig:subjective_measures}
\end{figure}

The quantitative results were analyzed across three dimensions: Technology Acceptance, Functional Design, and Parental Experience.

\textbf{Technology Acceptance (TAM): }
No significant differences were observed between the two systems in terms of perceived ease of use (p = 0.15), helpfulness (p = 0.41), or intention to use in the future (p = 0.56). The slight variation in ease of use may be attributed to differences in the voice recognition experience. As one parent noted, for younger children with less clear articulation, Doubao allows users to type in corrections, whereas PAPEL's voice-only interaction requires users to speak more clearly. Overall, parents gave consistently positive evaluations for both Doubao and PAPEL.

\textbf{Functional Design: }
Significant improvements were observed in content organization (p = 0.01), balancing entertainment with educational value (p = 0.03), and expression guidance (p = 0.03), indicating more effective support for language learning through structured and engaging interactions. No significant difference was found in vocabulary selection (p = 0.63), as parents considered both Doubao and the system effective, despite differences in their delivery formats. In both experiments, parents' word choice largely relied on intuition, with both systems being used as translation tools. A more detailed analysis and summary of the functional support were further elaborated in the interviews, particularly concerning how the system facilitated parental involvement.

\textbf{Parental Experience: }
Effort increased significantly (p = 0.04), as some parents noted that the system offered more structured input, prompting them to engage in deeper reflection and provide more intentional support. This suggests that the system placed higher demands on parental involvement, contributing to the observed increase in effort. While this may appear burdensome, we consider it a potentially positive outcome, as it encourages parents to think more critically about their instructional choices and become more actively engaged in the learning process. No significant changes were found in self-efficacy (p = 0.29) or frustration (p = 0.38); however, these two measures varied notably across participants. For example, some parents reported feeling overwhelmed by information while using PAPEL, whereas they experienced smoother interactions with the more familiar Doubao, despite its different organizational structure. Overall, most parents reported feeling capable and encountering few difficulties when using both systems. 

\subsection{How PAPEL Supports Parents in Play-based English Learning}
\label{subsec:interview}
In the previous two sections, we used quantitative measures to provide an initial view of PAPEL's positive impact on the play-based English learning process, as well as parents' experiences using the system, accompanied by preliminary interpretations. In this section, we draw on interview data and video data to provide a more in-depth explanation of how PAPEL supports parents in this context, further enabling children to achieve enhanced learning outcomes while preserving play's inherent fun and exploration.

\subsubsection{\textbf{Providing Personalized and Contextualized Content Support}}
PAPEL enhances parental instructional engagement by embedding target vocabulary within rich, play-based scenarios. Beyond offering isolated words or simple translations, it generates complete sentences paired with relevant play cues, helping parents integrate language teaching into interactive contexts. According to P7: \textit{``When I told him `watermelon', it gave me the sentence `I like juicy watermelon,' and then I read it to my child.''}  P1 similarly recalled: \textit{``For example, when I ask how to say `brush teeth,' it connects to a sentence like `I want to brush my teeth every day.' I think this kind of association is very helpful.''} These outputs support both language modeling and meaningful parent-child interaction, ensuring continuous learning input throughout children's play experience.

The system also personalizes content to parents' interests and habits, broadening the scope of both instruction and play. Multiple parents (P2, P5-8, P11-13) reported that the AI can provide a basic scene and adjust content based on the dyads' own interests. This flexibility empowers parents to decide what to teach and how to embed learning into daily routines, offering a seamless fusion of teaching and play.

In play-based settings where teaching paths are often unclear, the system offers structural and approporiate cues that guide both talk and activity. As expressed by P5 and P15: \textit{``It can help me design a basic scenario, and then I just follow its guidance and I don't need to plan or prepare anything.''} Compared to dictionary-style tools, the system was seen as a more proactive partner in both instruction and play. P6 described: \textit{``Within a fixed scene, I had to keep thinking and force myself to teach, but with such a guide, it lightens that pressure and even improves my mood.''} Beyond content delivery, multiple parents (P1, P6-P8, P11) reported that contextualized content support stimulated creative thinking by enriching the scene.

Despite these strengths, limitations remain. The system lacks fine-tuned adaptation to children's comprehension levels, placing a filtering burden on parents. As P8 pointed out: \textit{``You have to choose words your child can understand… and I don't even know how to say many words myself.''} Additionally, real-time guidance can be hard to follow during play. Parents expressed a need for clearer pre-session planning. As P12 suggested: \textit{``There should be a preparation phase beforehand; it would be better if the AI could provide a video or outline.''}

These findings highlight the value of an AI that provides both teaching and play content, while underscoring the importance of balancing in-the-moment guidance with structured, preparatory support.


\subsubsection{\textbf{Adaptive Language Mediation: Supporting Triadic Interaction Through Context-Aware Communication}}
PAPEL dynamically adjusts its output to match the English proficiency of both parents and children, thereby enhancing parents' instructional capacity. Many parents adopted a mediator role, paraphrasing or simplifying AI-generated expressions based on their child's reactions. As P13 shared: \textit{``I would take the child to restate what the AI said.''} This triadic interaction model enabled real-time adaptation and improved modeling efficacy.

PAPEL also lowered the barrier for participation, making it easier for parents to repeat and model language confidently. Reflecting on this, P10 remarked: \textit{``The term for `intravenous drip' was relatively long---I didn't know it at all. But I just repeated what the AI system said. This makes it easier for me to get involved.''} P8 further stated: \textit{``Even with enough vocabulary, lack of usage leads to unidiomatic speech and fear of mistakes. Having PAPEL provides helpful prompts, and I become more confident to speak.''}

For children, PAPEL's use of age-appropriate language aligned well with their expressive tendencies, encouraging imitation and engagement.
P1 pointed out: \textit{``When asking `how do you say spoon,' it might repeat it, enhancing my kid's memory, and my kid is willing to repeat it because she's familiar with the repeating behavior in her own English class.''} At the expressive level, we also found that this adaptive language use successfully guided children to more actively adopt a code-switching language pattern. The cases of D4, D11 and D6 are quite similar: At first, they could only repeat words, but later, after hearing PAPEL's expressions multiple times, they also began to naturally embed English words into their sentences.

Yet some limitations remain. P5 mentioned: \textit{``When I ask it in English, I can't fully understand its response, yet still need to repeat to my child.''} P12 also noted: \textit{``Young children use abbreviations like `veges', but both Doubao and PAPEL can't handle these expressions well.''} Instances of children mimicking awkward or inaccurate expressions further underscored the need for improved linguistic calibration and age-appropriate design.

\subsubsection{\textbf{Balancing Playfulness and Pedagogy: Supporting Naturalistic Language Learning Through Embedded Interaction}}
PAPEL's contextualized design reduced parents' teaching deliberateness and pressure while enhancing children's initiative and activating parents' awareness of how to balance learning and play. Parents appreciated the preparation-free guidance, with P3 and P16 similarly noting that ``it lightens the workload.'' This approach created a relaxed atmosphere, as P6 described: \textit{``My child obviously felt it was completely a playing process; he's enjoying it without feeling any pressure.''} P14 observed: \textit{``My child thinks it (PAPEL) is another Doubao (baseline) at first. But soon he thinks he's playing. \ldots During the break, he told me he enjoyed the session and happily repeated the words he learned from PAPEL many times, and after the second session he just looked tired.''} P8 also emphasized how contextual scenarios enhanced children's acceptance and natural immersion in activities. This shift from explicit teaching to contextualized play fostered genuine engagement for both parents and children.

On the other hand, the cases of D10 and D14 illustrate how PAPEL intervened when interactions became overly playful. Because P10 and P14 had limited English proficiency, they tended to focus on playing with the child and struggled to incorporate instructional guidance. During their interactions with PAPEL, the system embedded simple target words in its responses and suggested concrete ways for parents to prompt the child to use them. With PAPEL’s guidance, P10 began to use basic English words herself, while P14—despite her low proficiency—enhanced the instructional value of the activity by having the child follow and respond to PAPEL’s responses.

In summary, PAPEL's contextualized design effectively supported parents in balancing playfulness with education. By reducing teaching deliberateness through scenario-based guidance, it fostered parent-child consensus on play-based learning. These findings suggest that PAPEL can help parents introduce instructional opportunities while maintaining a playful interactional flow.

\subsubsection{\textbf{Semi-proactive Guidance as Scaffolding: Tailoring Support to Varying Parental Roles and Needs}}
PAPEL's proactive guidance provided timely support for parents when the process began to drift or stall. As P9 noted: \textit{``I'm not a professional educator, so I tend to be disorganized and just say whatever comes to mind. \ldots PAPEL sometimes set a new scenario proactively when I'm stuck; I feel like it's more scientific and helpful.''} Echoing this, P6 remarked: \textit{``I'm not good at this; I might just say whatever I see.\ldots The AI’s suggestions can lead us to a new activity, making our play more fun and natural.''} Through timely extra guidance, the system relieved parents from the pressure of content design, allowing them to ``follow along'' with minimal preparation.  

Meanwhile, we found typical cases in session recording videos showing the benefits of proactive guidance for children. During D16’s playing session, there was a period when the parent kept introducing new words, but the child was only repeating them verbally while playing with their own toys. When the parent asked about the English word for candied hawthorn, PAPEL shifted the topic to the pot in the child’s hands, starting a play scenario of making candied hawthorn, which immediately captured the child’s attention.

More proactive and instructionally savvy parents demonstrated bidirectional guidance with the system. P8 said that instead of dictating the play, PAPEL acted as a creative catalyst, providing just enough structure without taking over. P14 described a clearer scene:
 \textit{``We were playing doctor, but my son just kept using the toy stethoscope over and over. I just asked simply about what to do next, but unexpectedly it (PAPEL) gave a suggestion: `Maybe the teddy bear has a tummy ache from eating too much honey!' It was the perfect little nudge.''} This semi-proactive mechanism can adapt to different levels of parental involvement in open-ended play, supporting deeper, more personalized interactions.

However, this flexibility also revealed certain challenges and tensions. Reflecting on her experience, one concern involved the pacing of AI responses, which sometimes disrupted the parent-child dialogue, as P5 explained: \textit{``The AI's suggestions actually interrupted my conversation with the child. It would say a whole chunk, and by the time I listened, my child was no longer paying attention.''} Beyond this, some parents also articulated higher aspirations for the AI's role. As P16 proposed: \textit{``The AI could act as an expert or teacher. Based on the interaction between me and my child, it could reason and supplement the teaching accordingly---as a kind of expert system.''} This reflects an expectation that proactive guidance should go beyond task-level assistance, offering strategic support in planning, context interpretation, and intent recognition to foster deeper collaboration.

\section{Discussion}

This discussion follows current theoretical streams in in-home learning and proceeds in two steps. First, we draw on Role Theory~\cite{role_theory} and our empirical findings to argue that AI for in-home learning should explicitly balance parents’ multiple roles. We then zoom in on the educator side of this balance, using the Pedagogical Play Framework to unpack four key functions—design, guidance, modeling, and reflection—and examine how AI can share these tasks with parents so that educator work is woven into play without undermining their natural strengths as playmates.

\subsection{Beyond Supporting Parental Roles: Balancing Parents’ Multiple Roles in In-Home Learning}

Our findings in Section~\ref{subsec:quantitative analysis} highlight a tension at the heart of parents’ dual role as playmate and educator. When parents tried to act like a teacher and a playmate at the same time, they led tightly structured question–answer routines and treated English as a discrete lesson inserted into play; children's learning experience improved little and was sometimes even hindered. In contrast, dyads who naturally ``acted as playmates'' achieved substantial improvement.  

Role Theory~\cite{role_theory} helps explain why such different orientations mattered. From this perspective, “playmate” and “educator” are distinct social roles, each tied to a culturally shared set of expectations about how adults should speak, act, and organize interaction with children. Role Conflict~\cite{role_conflict} arises when the expectations of parents' two roles are activated in the same situation but point in different directions. This conflict is also relational: preschoolers at home naturally gravitate toward free play; as a result, when parents try to enact a dual playmate–educator role, they often face resistance or breakdowns that make the role conflict salient.
A deeper reason lies in in-home learning itself: parents expect to handle both playmate and educator roles, yet while the playmate role comes more naturally, the educator role is less familiar and more demanding---one they may feel less capable of, but still aspire to take on. 

Existing work about AI for in-home learning has already recognized that parents occupy multiple roles and has provided corresponding support~\cite{yu2024parent,yu2021parental_mediation,yu2021parental_facilitation}. In structured activities such as co-reading~\cite{Social_robots_as_conversational_catalysts,xu2023rosita} and storytelling~\cite{zhang2022storybuddy,Storytelling}, prior systems often emphasize parent-led interaction by generating and delivering appropriate content~\cite{liu2025bricksmart,zhang2022storybuddy} and by taking on different roles themselves~\cite{10.1145/3706598.3713330,ho2024s,Social_robots_as_conversational_catalysts}. In these designs, decisions about when and how parents shift among their multiple roles are left largely to their own judgment~\cite{10.1145/3706598.3713405,10.1145/3706598.3713330}.
However, in more open-ended settings like play-based learning, the child-led, fast-paced interaction demands that parents simultaneously inhabit playmate and educator roles, placing heavy demands on their ability to manage and coordinate both.
In East Asian cultures where academic performance and behavioral norms are strongly valued, the pull of the educator role is especially salient, making parents particularly prone to copying classroom scripts into home play~\cite{10.1145/3479543}. Neglecting this issue may undermine the quality of parent–child shared experience and even the family relationship itself~\cite{yang2025familycommunication}.

Within a parent-centric research paradigm, in light of role tensions of parents and children, we therefore argue that across in-home learning activities, AI systems should not only support the various roles parents may occupy, but also actively mediate and rebalance them in ways that are sensitive to the activity at hand. In more open-ended settings like play-based learning, this general principle takes a specific form: on one hand, systems should provide enough instructional scaffolding~\cite{zhang2022storybuddy,kwon2022captivate} to ease parents’ real-time decision-making effort; on the other hand, through AI-driven story progression and prompts, systems should gently steer parents back toward a playmate-centered stance that still incorporates light educator moves, so that learning goals are pursued indirectly through smooth, engaging play. 

PAPEL is designed not only to accommodate parents’ dual roles as playmate and educator, but to actively balance them during open-ended play. It does so by explicitly modeling a balance state, allowing the system to detect moments where one dimension appears less emphasized and gentle support may be helpful. Based on this state, PAPEL provides optional, extendable responses as light-touch nudges that preserve the flow of play and keep control with the parent. Our user study suggests that this balancing mechanism can strengthen parents’ facilitation in play-based learning, while also surfacing the underlying tension.
Future research on AI systems for in-home learning could also treat the balance of parents’ multiple roles as a first-class design goal and further operationalize and refine role-balance designs to achieve stronger outcomes. This framing positions role balance---not just content quality or task completion---as a central success criterion for AI support. 

\subsection{AI as Pedagogical Partner: Supporting Parents’ Educator Role in Play-Based Learning}
\label{subsec:pedagogical play discussion}
The previous discussion highlights the importance of balancing parents’ multiple roles, yet it also leaves open a crucial problem: both the parent and child are generally comfortable engaging as playmates, but parents often struggle to sustain a dual playmate–educator role in ways that also elicit a corresponding “player–learner” stance from the child. In contrast, trained teachers can more reliably orchestrate this dual-role alignment in the moment. Many parents nonetheless hope to perform like teachers during in-home activities, yet parents are not trained teachers, and effective in-home learning cannot rely on simply imitating classroom practice. Therefore, in this section, we focus on how AI can scaffold parents’ educator role rather than turning them into teachers.

Specifically, we turn to the \textit{Pedagogical Play Framework}~\cite{pedagogical_play}, which builds on three well-established types of play: open-ended play, modeled play, and purposely framed play~\cite{pedagogical-play-types}. We identified four key concepts in the framework, which were also frequently mentioned in the post-user study interviews: design, the planning of learning activities (from purposely framed play); guidance, moment-to-moment support for children’s exploration (from modeled play); modeling, the in-situ demonstration and rehearsal of target skills (from modeled play); and reflection, the observation of and adaptation to children’s needs in and over time (from open-ended play).

These four concepts provide the lens through which we analyze how AI can help parents weave educator-like work into play with children while staying in a playmate role. In the following sections, we discuss how our system design aligns with these pedagogical principles and outline theory-informed research spaces and design implications grounded in our empirical findings.

\subsubsection{Design: AI as a Co-planner of Play-based Learning}
In educational play, activity design sets the foundation for engagement and learning \cite{guz2016learning}. While teachers often draw on curriculum knowledge and developmental goals to structure meaningful scenarios, many parents lack the pedagogical expertise to take on this full role at home. AI has the potential to bridge this gap by supporting families in transforming abstract learning goals into concrete instructional content \cite{liu2025bricksmart,zhang2022storybuddy}, dynamically adjusting based on context \cite{mathemyths,ho2024s}, and offering personalized content design \cite{KF,lee2024open}. Our work aligns with these objectives by facilitating real-time content generation based on play context while encouraging more active educational input.

Meanwhile, our findings reveal that parents often desire more than just content generation -- they want end-to-end support in designing learning activities that fit their child and daily routines. In interviews, several parents asked not only for concrete instructional frameworks, but also for help in turning their deep knowledge of their children into tailored play-based scenarios. In this view, the design phase becomes the place where parents think together with AI: parents contribute goals, constraints, and personal observations of their child, while AI proposes and refines activity structures, step sequences, and ways of embedding target language into ordinary play. Once the activity starts, parents can focus on maintaining a playful, responsive interaction, letting learning emerge naturally, rather than constantly switching into an educator role.

\subsubsection{Guidance: AI as Conflict-solver within Triadic Interaction}
Guidance in pedagogical play requires adults to support children’s exploration while resisting the urge to overly direct. Much existing work on AI-supported in-home learning has examined relatively structured activities with fixed topics or preset trajectories, such as specific books, stories, or tasks \cite{liu2025bricksmart,zhang2022storybuddy,xu2023rosita}. In these settings, systems can anticipate likely next steps and offer parents concrete prompts or questions that help them guide the interaction along a known path. Such support is effective when the learning goal and activity frame are clear in advance.

In open-ended play, however, topics shift rapidly and goals are only loosely specified, which makes guidance harder. Our findings show that parents can usually provide moderate guidance on their own, but struggle at unexpected turns in play---when a child suddenly shifts the storyline, the parent anticipates something else, and the parent is torn between following the child or pulling the activity back toward the presumed goal. These divergence moments are where role tensions surface most strongly. Parents who are unsure how to proceed tend to slip into a more controlling role, attempting to “get things back on track”, which gradually amplifies role conflict.

This suggests a different emphasis for AI in open-ended guidance. Instead of only delivering a stream of content for parents to say or do, AI should help parents navigate these moments of confusion and divergence with their children. In role terms, AI takes on the work of making sense of an unpredictable interaction and organizing possible ways forward, so that parents can stay primarily in a playmate stance, choosing among options and offering light, situational guidance instead of feeling compelled to reassert a full teacher role whenever play drifts away from the preset plan.

\subsubsection{Modeling: AI as Knowledge Transmission Scaffold}
In modeled play, teachers introduce new language through purposeful demonstration and repetition. At home, this educator role is harder for many parents to take up: varying English proficiency levels and low teaching confidence make it difficult for them to see themselves as competent language models. As noted above, most existing research has emphasized the provision of instructional content \cite{zhang2022storybuddy,xu2023rosita,kwon2022captivate}, with relatively little attention to how parents can effectively deliver that content \cite{liu2025bricksmart}. 

In PAPEL, we instead observe modeling emerging through imitation of the AI’s language use, especially its code-switching strategy. The system typically blends the parent’s first language with simple English phrases or keywords, embedding target language in a familiar linguistic frame. Many parents spontaneously mimicked this pattern and gradually extended it with their own ideas, which is ideal for children's learning. In effect, AI-generated utterances functioned as reusable “language templates”: parents could repeat and adapt them without needing to invent pedagogical English from scratch. This AI-scaffolded imitation lowered the barrier to modeling, allowing parents to project their own intentions and examples through a form they felt able to perform, and to step into a light educator role while still behaving like a playful, responsive partner. In role terms, AI provides stable, easy-to-copy linguistic scaffolds, while parents remain the embodied models in front of the child, delivering language in their own voice and within ongoing play.

Language learning presents a unique advantage here, as language functions both as knowledge and as the medium of delivery, making it relatively straightforward to weave modeled language into play. In other domains such as science or math, this seamlessness may not hold; future systems will need to explore how AI can still support modeling by suggesting simple demonstrations and activating parents’ own domain knowledge, while keeping parents in the foreground as relational, context-sensitive demonstrators.

\subsubsection{Reflection: AI as Awareness Promoter}
Reflection in play involves observing, interpreting, and adjusting interaction strategies. Earlier, we argued that AI for in-home learning should help balance parents’ multiple roles. From a pedagogical perspective, reflection is the process that links this role balance to learning: over time, adults learn how different role choices lead to different interactional and learning outcomes. For professional educators, this kind of reflective practice is part of their training; for parents, especially in fast-paced play-based settings, it is rarely explicit and is hard to sustain in the moment. 

Some existing work focuses on helping or guiding parents to reflect on and adjust their instructional approaches, including supporting reflective adaptation processes \cite{dietz2024contextq,zhang2022storybuddy,liu2025bricksmart} or encouraging deeper consideration of teaching practices \cite{dietz2024contextq,xu2023rosita,ho2023designing}. In our study, beyond simply adjusting content, parents were also required to adapt to an implicit goal: achieving a balance between play and learning. Consistent with prior research, our system does not require parents to engage in such reflective thinking or adjustments themselves but instead directly assists them in achieving the desired balance.
However, our findings suggest that while reducing cognitive load, such a reflection design also means that parents’ awareness of their own role patterns remains limited.

This suggests a clear role for future AI: instead of asking parents to self-monitor and evaluate every interaction, the system can take on much of the evaluator work and offer light-touch interventions. Concretely, AI can track patterns in how parents move between playmate and educator roles, relate these patterns to signals of engagement or learning, and surface brief, contextual feedback. The goal is not to judge parents, but to make the link between their role patterns and interactional effects more visible, so that parents can decide for themselves, based on their knowledge of the child and family values, whether and how to adjust their stance. This suggests a broader design space for role-aware reflective support: future systems can make the impact of different role choices more visible across sessions, while leaving parents in control of how to respond. Over time, this can help parents develop a more flexible, transferable awareness of how their roles shape play-based learning at home.

\subsection{Limitation and Future Work}

This study has several limitations. The sample size was relatively small and predominantly comprised families from middle to upper-middle-class backgrounds in China, with higher-than-average educational attainment and awareness of early education. In addition, there was a gender imbalance among the child participants, which may have influenced the findings. As such, the findings may not be generalizable to families from different cultural or socioeconomic backgrounds. Future studies should engage diverse populations to explore how AI-supported play-based learning systems can be adapted to different household contexts.

In terms of research focus, this study concentrates on how AI systems influence parents’ educational practices and perspectives, with comparatively less attention to children’s learning outcomes. Future work could build on more playful methods to collect children’s feedback \cite{playful_method_for_feedback_collection}, thereby gaining richer insights from the children’s perspective.

We note that our current performance modeling remains relatively simple and heuristic-based, serving primarily as a lightweight signal for guiding system behavior, not a comprehensive measure of parent--child interaction quality. Future work should also more tightly connect interaction modeling and intervention design. In particular, researchers could develop more fine-grained and validated measures of parent--child performance from multimodal interaction data, and then use these signals to drive more principled response strategies. Such an approach would allow future systems to better decide when to intervene, what type of support to provide, and how to balance playful and instructional goals, thereby enabling more transparent and systematically controllable AI behavior in open-ended parent--child learning scenarios.

In addition, the baseline in this study reflected the general-purpose web-based chatbots that were practically accessible at the time of our system design and experiment. As conversational AI continues to evolve toward more interactive and multimodal forms, future studies could revisit parent-child play-based learning with newer multimodal-mediated chatbots. Comparing PAPEL-like support with these emerging tools may reveal new patterns of parent adaptation, child engagement, and parent--child--AI interaction that were not yet observable in the present study.

From a technical perspective, current VLMs exhibited limited capability in recognizing toys and play objects, which constrained their ability to support nuanced context understanding during play. Future work could explore more robust perception models tailored to open-ended play scenarios to improve system responsiveness and relevance. Additionally, the fragmented and often unclear speech patterns typical of young children posed challenges for speech recognition systems, affecting the system’s performance. Improving robustness in child speech recognition remains an important technical direction to enhance the overall user experience and accuracy of AI assistance in family play settings.

\section{Conclusion}
We present PAPEL, a parent-AI collaborative system designed to support preschoolers' play-based English learning. Grounded in a two-phase formative study, we identified key challenges faced by parents and explored how AI can help navigate these challenges, leading to four general principles for designing AI systems in play-based learning contexts. Guided by these principles, we developed PAPEL and conducted a mixed-method user study with 16 parent-child dyads. Our findings suggest that PAPEL enhanced parents' instructional proficiency, increased the overall engagement of interactions, and improved parents' ability to integrate educational content with playful experiences. We close by arguing that AI should act as a role-aware partner that actively rebalances parents' playmate and educator roles, offering new design insights for future family-centered AI systems.

\begin{acks}
This work was supported by the Natural Science Foundation of China under Grant No. 62132010. We acknowledge the use of LLMs as writing assistants for language polishing and term translation. All AI-generated suggestions were reviewed and edited by the authors, who take full responsibility for the final content.
\end{acks}

\bibliographystyle{ACM-Reference-Format}
\bibliography{main}


\appendix
\newpage
\section{Demographic Details of Participants in the Formative Study}
Edu.= parent's education; Eng. Prof.=self-reported English proficiency (1–5); B.A./B.S.=Bachelor’s, M.A./M.S.=Master’s, Ph.D.=Doctoral; M/F=male/female; 
child English exposure is approximated by time spent in formal English classes. Parent English proficiency is self-reported on a 1–5 scale (1 = minimal ability to communicate in English; 5 = frequent use of English at work and no difficulty in everyday conversation).

\begin{table}[htbp]
\centering
\begin{tabular}{lccccc}
\toprule
ID & Edu. & \makecell{Parent Eng.\\Prof. (1--5)} &
\makecell{Child\\Age} & \makecell{Child English\\Learning (months)} &
\makecell{Prior\\In-home Teaching} \\
\midrule
P1 & B.A. & 2 & 3 & -    & Yes \\
P2 & B.A.  & 4 & 3 & 3    & Yes \\
P3 & M.A.    & 3 & 4 & -    & Yes \\
P4 & B.A.  & 3 & 6 & 7    & Yes \\
P5 & Ph.D. & 5 & 6 & 36+  & Yes \\
P6 & B.A.  & 3 & 4 & -    & Yes \\
P7 & M.A.    & 2 & 5 & 12   & Yes \\
P8 & M.A.    & 4 & 4 & 16   & Yes \\

\bottomrule
\end{tabular}
\caption{Information about interviewed parents.} 
\label{tab:child-info-experience}
\end{table}

\begin{table}[htbp]
\centering
\begin{tabular}{cccc}
\toprule
ID & Workplace & Teaching Experience (yrs) & Gender \\
\midrule
T1 & International Kindergarten & 7 & M \\
T2 & After-school Education Center & 8 & F \\
\bottomrule
\end{tabular}
\caption{Information about teachers in the interview.}
\label{tab:teacher-info}
\end{table}

\begin{table}[htbp]
\centering
\begin{tabular}{lccccccc}
\toprule
ID &
\makecell{Edu.} &
\makecell{Parent Eng.\\Prof. (1--5)} &
\makecell{Parent\\Gender} &
\makecell{Child\\Age} &
\makecell{Child English\\Learning (months)} &
\makecell{Child\\Gender} &
\makecell{Prior In-home\\Teaching} \\
\midrule
D1 & B.A. & 1 & F & 3 & 3  & M & No  \\
D2 & B.A. & 3 & F & 4 & 6  & F & Yes \\
D3 & B.A. & 2 & F & 6 & 12 & M & No  \\
D4 & M.A. & 4 & F & 4 & 6  & F & Yes \\
D5 & M.A. & 3 & M & 5 & 2  & F & No  \\
\bottomrule
\end{tabular}
\caption{Information about parent-child dyads in the Wizard-of-Oz study.}
\label{tab:child-parent-info}
\end{table}

\section{System Detail}
In this section, we introduce the detailed design and implementation of PAPEL.

\subsection{Content Generation}
\label{appendix:implementation-content-gen}
For vocabulary recommendations, we curated a base vocabulary set of 100 words tailored to our play scenarios and informed by teacher interviews, which identified common and developmentally appropriate lexical categories for preschool learners. Building on this resource, an LLM-based vocabulary module automatically selects five candidate words or generates new ones solely based on the current interaction context.

For game guidance, we distinguish between two types of prompts identified in teacher interviews: narrative prompts, which continue and deepen the current storyline, and interactive prompts, which enrich and extend it through structured activities. For narrative prompts, the LLM directly generates three suggestions conditioned on the current scene and recent dialogue. For interactive prompts, we first derived a small library of five reusable activity templates from teacher and parent interviews (e.g., hiding one item and asking the child in English to identify what is missing; guessing an object or character from its function; following English action commands while emphasizing target words; sorting objects by attributes and describing them in English; and role-play style situational dialogues). The LLM then contextualizes these templates using the ongoing conversation history to produce concrete, scenario-appropriate suggestions. To reduce latency, we employ a chain-of-thought prompting scheme that performs both narrative prompt generation and interactive prompt contextualization within a single LLM call.

The parallel design for content generation helps us better preserve the different functional goals of the two prompt types and leads to more stable and appropriate guidance outputs.

\subsection{Language Adaptation}
\label{appendix:implementation-language}

To adapt the system's output to the language proficiency of both the parent and the child, the system performs lightweight analysis on recent dialogue. Drawing on code-switching theory\cite{mannava2024exploring}, language use in bilingual interactions can be categorized into two primary patterns: word-level mixing and sentence-level mixing. Building on this framework, and in collaboration with an English teacher, we defined three proficiency levels and corresponding developmentally appropriate response strategies: (1) word-level mixing, (2) sentence-level mixing, and (3) basic conversational English.

Concretely, if we detect three or more consecutive exchanges in which both parent and child use only English, we treat the dyad as capable of basic English conversation (level 3). In this case, the system replies in English throughout, using short, natural utterances and occasionally introducing slightly richer structures while avoiding unnecessary use of the first language. If the dialogue contains no fully English turns but exhibits natural sentence-level mixing, the dyad is classified as level 2. Here, the system responds in full English for short, simple sentences, but uses sentence-level mixing for longer or more complex content to maintain comprehensibility. When the dialogue only shows isolated English words embedded in otherwise Chinese utterances, the dyad is classified as level 1, and the system primarily uses short, simple sentences with sentence-level mixing, repeatedly highlighting the target English words to support initial vocabulary acquisition. In summary, our language adaptation strategy deliberately sets the difficulty of system output slightly above the dyad’s current language use, in order to scaffold the child’s learning. This design aligns with the notion of the zone of proximal development (ZPD) \cite{ZPD}.

\subsection{Balance Assessment}
\label{appendix:implementation-balance}

To balance the dimensions of playfulness and knowledge-oriented instruction, the system first evaluates scores for both aspects based on the recent dialogue history. The playfulness score $s_p$ is derived from two components that directly mirror our two types of game guidance: the fluency of narrative progression, which corresponds to narrative prompts, and the degree of conversational divergence, which corresponds to interactive prompts, following prior teacher recommendations for classroom assessment. The LLM generates a narrative score ($s_n$) and a divergence score ($s_d$), both represented as binary variables (0 or 1, indicating whether the narrative progresses smoothly and whether the conversation is sufficiently varied). The total playfulness score is then calculated as the sum of these two components ($s_p = s_n + s_d$), resulting in a score ranging from 0 to 2 that provides a lightweight signal about the current play-oriented tendency of the interaction.

The educational score is based on the parent’s recent efforts to convey English knowledge, quantifying the extent and intensity of the language content they provide. In our prior Wizard-of-Oz study, we observed that parents’ instructional behaviors often became highly repetitive, prompting us to explicitly encode this pattern into the score design. The educational score, denoted as $s_e$, thus reflects how strongly the parent emphasizes teaching language concepts over the recent three-minute window. Concretely, $s_e = 0$ when the parent provides no instructional guidance; $s_e = 1$ when the parent offers instructional guidance that is meaningfully interwoven with the ongoing game process; and $s_e = 2$ when the parent delivers frequent or repetitive instructional moves, indicating a strong instructional focus. Like the playfulness score, $s_e$ ranges from 0 to 2, but here higher values represent greater pedagogical intensity rather than higher quality. The system uses the absolute difference between the two scores as a lightweight trigger for adaptation: when $|s_p - s_e| \geq 1$, subsequent guidance is adjusted to modestly strengthen the less-emphasized dimension of the interaction.

This balance assessment module is also implemented using a chain-of-thought prompting scheme, in which a single LLM call jointly infers both $s_p$ and $s_e$ from the shared multimodal context.

\subsection{Proactive Guide and Response Generation}
\label{appendix:implementation-response}
To support proactive yet parent-led interaction, we implemented an extended response mechanism in the system. As illustrated in Fig.~\ref{fig:pipeline}, beyond generating contextually relevant responses, the system also analyzes the pedagogical or playful orientation of each parent utterance directed toward the system, based on input categories identified in our prior Wizard-of-Oz study (teaching-oriented, play-oriented, and neutral). These classifications are then compared with the dialogue-level orientation inferred from the balance assessment module, which monitors the evolving balance between pedagogical and playful content over time.
When the parent's current input aligns with the overall interaction trend (e.g., both being highly playful or pedagogical), the system may append an extended response to modestly support the less-emphasized dimension. Otherwise, we consider the parent to be potentially aware of the trend and to be trying to keep both on track. Specifically, if pedagogical content is lacking, the extended response emphasizes randomly recommended vocabulary items and encourages active verbal engagement between parent and child. Conversely, if the dialogue has become overly instructional, the system introduces playful elements inspired by the current game-step recommendations to re-engage the child in the activity.

In cases where the dialogue remains well-balanced, the system provides only a standard response. Meanwhile, vocabulary suggestions and game-step prompts are persistently displayed on the interface, serving as ongoing cues to help parents maintain attention to both instructional and playful dimensions of the activity.

In response generation, the LLM dynamically adjusts both the linguistic complexity and expressive style according to the strategy corresponding to the parent-child language level assessed by the previous module. Additionally, following expert recommendations, all responses to children are constructed using concise and simple English, while maintaining vivid and engaging expressions to promote sustained participation.

\subsection{Implementation}
\label{appendix:implementation}
The core system was implemented in Python, with a server developed using FastAPI \footnote{https://fastapi.tiangolo.com/} to provide RESTful API endpoints. All generative components relied on OpenAI's ChatCompletion API \footnote{https://platform.openai.com} to support large language model (LLM) inference. To balance latency and quality, different models were employed across tasks: for lightweight generation tasks such as vocabulary suggestions and play guidance (content generation module), the GPT-4o-mini model was adopted to ensure low response latency; for more complex tasks involving chain-of-thought prompting (language adaptation, balance assessment, and extended response modules), the GPT-4o model was used to ensure output reliability and coherence.

The client-side application was built as a desktop application using the Vue.js \footnote{https://vuejs.org} framework. The client communicates with the server via REST APIs, transmitting real-time audio recordings of parent–child interactions. Speech input was processed using the TINGWU speech recognition API \footnote{https://tingwu.aliyun.com}, which provides automatic transcription and speaker diarization. Given the challenges of accurately recognizing young children’s speech \cite{Kalanadhabhatta2024Playlogue}, additional post-processing was performed on the server using GPT-4o-mini to repair recognition errors and enhance transcript quality.

To further reduce perceived latency and ensure timely feedback, the system updates the outputs of core modules in the background at a fixed interval of every 15 seconds. This proactive update strategy ensures that when user input is received, the system can immediately generate a response with only one LLM call based on the most recent contextual information, resulting in an average response latency of approximately 3 seconds.

For output delivery, system responses were converted to speech using the ALIYUN streaming text-to-speech (TTS) API\footnote{https://ai.aliyun.com/nls/tts}, enabling real-time audio playback on the client device.

\section{Example Parent–Child Interaction with PAPEL}
\label{appendix:conversation}

This appendix presents a short excerpt of a parent–child interaction supported by PAPEL from D5. Original conversations involved Mandarin with English code-switching. We present transcripts in English; segments that were originally spoken in English are marked with angle brackets (e.g., <tooth>). System turns are prefixed with “system:”, and child and parent turns are lightly cleaned for readability. 

\begin{itshape}
\hspace*{1em}

parent: How about we play dentist now?

parent: Shall we check the little hippo's teeth? The little hippo is the patient now; he has come to see the doctor.

system: The little hippo is here! The English for "hippo" is <hippo>! Try calling it!

parent: Then you ask it, how do you say "hippo" in English?
child: <hippo>.

parent: <Are you feeling okay?> You come take care of this little hippo.

child: They just didn't grow very well. The black ones mean the teeth have decayed.

parent: What happened to the little hippo's teeth?

system: The little hippo's teeth are dark, so he may have cavities. We need to help him and check his teeth! <Let's check his teeth together.> We can start examining the little hippo's teeth now!

parent: There is a cavity there; that tooth is a <decayed tooth>. Ask again how to say "tooth".

child: How do you say it?

parent: <Let's together choose a toothbrush.> Which one is the toothbrush?

child: This one.

parent: Okay, <this is tooth brush.> Then you brush his teeth.

child: <toothbrush>.

parent: You need to brush his decayed tooth thoroughly.

parent: You have to brush every tooth clean, because brushing only one tooth is not enough.

parent: Right, so we must brush every tooth clean, because brushing just one tooth is not enough.

parent: The white ones too, right. You can ask the little hippo whether there are worms in his teeth.

child: I want to ask the little robot.

child: Little hippo, do your teeth have worms?

system: Great! You can ask the little hippo, "<What happened to your teeth?>" and pay attention to his cavities together! Children, we can use the English word for "cavity", which is "<cavity>", to ask about the little hippo's teeth. You can say "<Do you have a cavity?>".

child: "cavity" is <cavity>.

parent: <What's happened to cavity? Do you have a cavity?>

child: <do you have a cavity?>

parent: Tell me how many cavities he has, <how many cavities>. Count them and see how many cavities he has.

child: Four. I know that four is <four>.

parent: Then I'll ask you <how many teeth?>. Count how many teeth the little hippo has; <teeth> means teeth.

...

\end{itshape}

\section{Interview Outline}

1. Please briefly introduce your child's age and English learning background.

2. Have you used other AI chat tools or English learning tools to tutor your child? If so, what were your main purposes in using these tools?

3. In your opinion, does play-based learning help children learn English? If so, which aspects of language learning do you think it helps most?

4. Do you have experience guiding your child in English learning? Could you share how you typically do this?

5. During the baseline phase, how did you intentionally conduct English teaching through "playing house"? For example, did you actively design teaching tasks and activities within the game? Did AI features help with design?

6. During the experimental phase, how did you arrange games? Did you try using the prompts in the lower right corner, or did you more closely follow AI's direct guidance?

7. During the experimental phase, how did you conduct teaching? Did you refer to the prompts in the lower left corner, or focus mainly on the child's attention targets? Do you think AI helped you have clearer goals or plans in the teaching process? Was this change evident in the experimental phase?

8. Baseline: How did you guide your child to speak English, and how effective was it? Did you provide demonstrations? How did the child respond to demonstrations? Did AI play a demonstrative role?

9. Experiment: How did you guide your child to speak English, and how effective was it? Did you provide demonstrations? How did the child respond to demonstrations? Did AI help?

10. Do you think the demonstration process became smoother, more natural, and simpler?

11. In the baseline, how did you guide your child to learn English while playing (following the child's lead or arranging it yourself)? Specifically, how was it integrated with teaching? Did the baseline help with your guidance?

12. In the experiment, what changes occurred in your guidance approach? How did AI influence your guidance strategies? Specifically, how was it integrated with teaching? Did the experiment help with your guidance?

13. Did guidance become easier?
Reflection-Adjustment (corresponding to play-learning balance): In the baseline, what problems did you encounter (regarding process or your own performance, not vocabulary or steps)? How did you reflect and adjust with AI's help? In the experiment, what problems did you encounter? How did you reflect and adjust with AI's help? Do you think reflection and adjustment when using the system became more intuitive and simple, or did you find it smooth without problems?

14. Function Support: How effective were vocabulary recommendations? Were they used? 
How effective were game recommendations? Were they used? What help did AI replies provide? 
How did parents use this information? Did they tend to actively organize content or follow actively guided content? Under what circumstances did you use recommended content versus following actively guided content?

15. General Discussion: What differences do you see in AI's role between the two situations? What would be the ideal role? What kind of help should AI provide parents? Beyond the system, what other help could be provided? How to leverage strengths and compensate for weaknesses?

16. What other similar play-based learning activities do you have at home? How are they conducted? If so, how do you think AI could assist your existing practices?

\section{Subjective Measures}

1. The AI system is simple and easy to use. 

\textit{1 for strongly disagree, 7 for strongly agree}

2. The AI's functionality is helpful for my teaching. 

\textit{1 for strongly disagree, 7 for strongly agree}

3. I would continue to use this AI-assisted tool for interactive English teaching with my child. 

\textit{1 for strongly disagree, 7 for strongly agree}

4. AI helps me decide on appropriate teaching vocabulary puzzles. 

\textit{1 for strongly disagree, 7 for strongly agree}

5. AI helps me decide on how to organize teaching content for activities.

\textit{1 for strongly disagree, 7 for strongly agree}
 
6. AI can naturally help me balance game and teaching content, allowing
children to have fun and learn English. 

\textit{1 for strongly disagree, 7 for strongly agree}

7. AI can help me guide children to express themselves in English more effectively. 

\textit{1 for strongly disagree, 7 for strongly agree}

8. During the entire game and teaching process, I put in a lot of effort. 

\textit{1 for strongly disagree, 7 for strongly agree}

9. How satisfied are you overall with the AI-assisted parent-child interactive English teaching process itself? 

\textit{1 for very dissatisfied, 7 for very satisfied}

10. After using AI, I felt more confident and capable in teaching. 

\textit{1 for strongly disagree, 7 for strongly agree}

11. During the entire game and teaching process, I experienced a sense of frustration and defeat. 

\textit{1 for strongly disagree, 7 for strongly agree}







\end{document}